 \documentstyle[aas2pp4,epsf]{article}

\newcommand{\Ell}{E_\parallel}      
\newcommand{\rhoGJ}{\rho_{{\rm GJ}}}  
\newcommand{\Bp}{B_{\rm p}}        
\newcommand{\sgP}{\sigma_{\rm p}}  
\newcommand{\rlc}{\varpi_{\rm LC}} 
\newcommand{\Es}{\epsilon_{\rm s}} 
\newcommand{\inc}{\alpha_{\rm i}}  
\newcommand{\Pcv}{P_{\rm CV}}      
\newcommand{\etaICe}{\eta_{\rm IC}^{\rm e}}
\newcommand{\etaICg}{\eta_{\rm IC}^\gamma}

\newcommand{\etaP}{\eta_{\rm p}}
\lefthead{Hirotani, Harding, and Shibata}
\righthead{}
\begin{document}

\title{High-energy Emission from Pulsar Outer Magnetospheres}
\author{Kouichi Hirotani
        \footnote{Present address:
                  Max-Planck-Institut f$\ddot{\rm u}$r Kernphysik,
		  Postfach 103980, 
		  D-69029 Heidelberg, Germany}}
 \affil{Code 661.0, 
        Laboratory for High Energy Astrophysics,\\
        NASA/Goddard Space Flight Center, \\
        Greenbelt, MD~20771;\\
        Email: hirotani@mpi-hd.mpg.de\\
        \ \\
        {\rm To appear in the inaugural issue of
             {\it Progress in Astrophysics Researches}}}

\begin{abstract}
We investigate a stationary pair production cascade 
in the outer magnetosphere of an isolated, spinning neutron star.
The charge depletion due to global flows of charged particles,
causes a large electric field along the magnetic field lines.
Migratory electrons and/or positrons are accelerated by this field
to radiate gamma-rays via curvature and inverse-Compton 
processes.
Some of such gamma-rays collide with the X-rays 
to materialize as pairs in the gap.
The replenished charges partially screen the electric field, 
which is self-consistently solved 
together with the energy distribution of particles and gamma-rays
at each point along the field lines.
By solving the set of Maxwell and Boltzmann equations,
we demonstrate that an external injection of charged particles
at nearly Goldreich-Julian rate does not quench the gap 
but shifts its position
and that the particle energy distribution cannot be described
by a power-law.
The injected particles are accelerated in the gap
and escape from it with large Lorentz factors.
We show that such escaping particles migrating outside of the gap
contribute significantly to the gamma-ray luminosity for
young pulsars
and that the soft gamma-ray spectrum between 100~MeV and 3~GeV
observed for the Vela pulsar can be explained by this component.
We also discuss that the luminosity of the gamma-rays
emitted by the escaping particles 
is naturally proportional to 
the square root of the spin-down luminosity.
\end{abstract}

\keywords{gamma-rays: observations 
       -- gamma-rays: theory 
       -- magnetic fields 
       -- methods: numerical
       -- pulsars: individual 
          (Geminga pulsar, PSR~B1055-52, PSR~B1706-44, Vela pulsar)
         }


\section{Introduction}
\label{sec:intro}
Recent years have seen a renewal of interest in the
theory of particle acceleration in pulsar magnetospheres,
after the launch of the {\it Compton Gamma-ray Observatory} (CGRO).
The Energetic Gamma Ray Experiment Telescope (EGRET) on board the CGRO
has detected pulsed signals from at least seven
rotation-powered pulsars
(for the Crab pulsar, Nolan et al. 1993, Fierro et al. 1998;
 for the Vela pulsar, Kanbach et al. 1994, Fierro et al. 1998;
 for PSR~B1706-44, Thompson et al. 1996;
 for PSR~B1951, Ramanamurthy et al. 1995;
 for PSR~B1046--58, Kaspi at al. 2000;
 for Geminga, Mayer-Hasselwander et al. 1994, Fierro et al. 1998; 
 for PSR~B1055-52, Thompson et al. 1999).
Since interpreting $\gamma$-rays should be less ambiguous
compared with reprocessed, non-thermal X-rays,
the $\gamma$-ray pulsations observed from these objects
are particularly important as a direct signature of 
basic non-thermal processes in pulsar magnetospheres,
and potentially should help to discriminate among different emission
models.

\begin{figure} 
\centerline{ \epsfxsize=8cm \epsfbox[0 0 420 200]{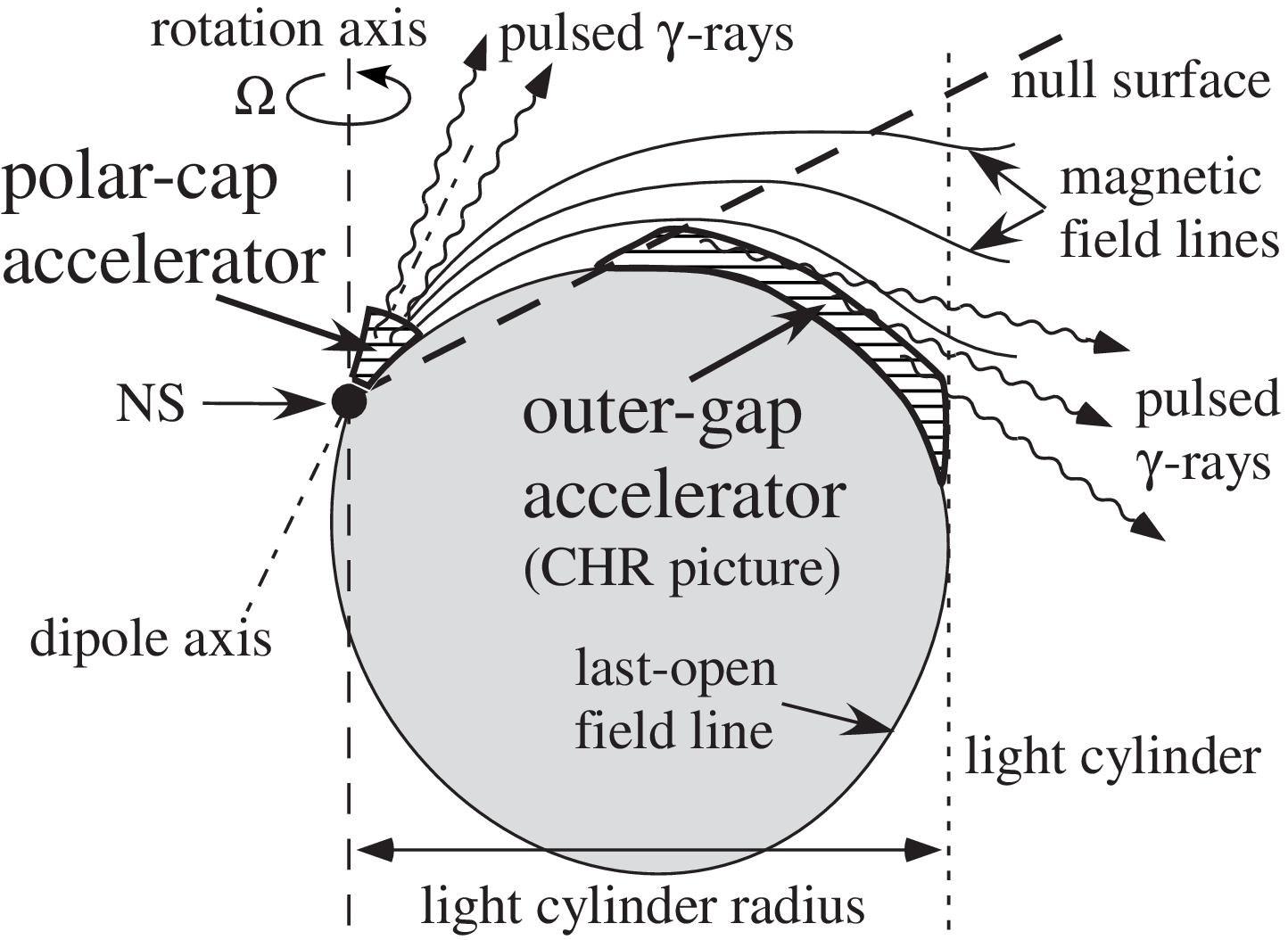} } 
\caption{
Schematic picture (side view) of the two representative models 
for accelerators (hatched regions) in a pulsar magnetosphere.
The small filled circle indicated by \lq NS' represents the
neutron star, which is rotating around the rotation axis
(light dashed line).
The light cylinder (dotted line) is located typically
a few hundred neutron-star radii from the rotation axis. 
On the null surface (heavy dashed line),
the magnetic field component projected along the rotation axis, vanishes.
The closed field lines, which do not penetrate the light cylinder,
are in the shaded region.
It is hypothesized in the CHR picture 
that the outer-gap accelerator extends
between the null surface and the light cylinder.
        }
\label{fig:sideview}
\end{figure}

Attempts to model the pulsed $\gamma$-ray emissions
have concentrated on two scenarios (fig.~\ref{fig:sideview}):
Polar cap models with emission altitudes of $\sim 10^4$cm
to several neutron star radii over a pulsar polar cap surface 
(Harding, Tademaru, \& Esposito 1978; Daugherty \& Harding 1982, 1996;
 Dermer \& Sturner 1994; Sturner, Dermer, \& Michel 1995;
 also see Scharlemann, Arons, \& Fawley 1978 for the slot gap model),
and outer gap models with acceleration occurring in the open
field zone located near the light cylinder
(Cheng, Ho, \& Ruderman 1986a,b, hereafter CHRa,b;
 Chiang \& Romani 1992, 1994; Romani and Yadigaroglu 1995; Romani 1996).
Both models predict that electrons and positrons are
accelerated in a charge depletion region, a potential gap,
by the electric field along the magnetic field lines
to radiate high-energy $\gamma$-rays via the curvature process.
However, there is an important difference between these two models:
An polar-gap accelerator releases very little angular momenta,
while an outer-gap one could radiate them efficiently.
In addition, three-dimensional outer-gap models
commonly explain double-peak light curves with strong bridges
observed for the $\gamma$-ray pulsars.
The purpose of the present paper is, therefore, 
to explore further into the analysis of the outer-gap models.

If a magnetized neutron star is rotating with angular velocity $\Omega$,
a static observer measures the rotationally induced charge density
(Goldreich \& Julian~1969; Mestel~1971)
\begin{equation}
  \rhoGJ 
  \equiv -\frac{1}{4\pi c}
         \nabla \cdot \left[ (      \mbox{\boldmath$\Omega$}
                             \times \mbox{\boldmath$r$}     )
                             \times \mbox{\boldmath$B$}      \right],
  \label{eq:def_rhoGJ_0}
\end{equation}
where $\mbox{\boldmath$\Omega$}$ satisfies 
$\vert\mbox{\boldmath$\Omega$}\vert=\Omega$ and 
directs the rotation axis (fig.~\ref{fig:sideview}),
$\mbox{\boldmath$B$}$ is the magnetic field at potision
$\mbox{\boldmath$r$}$ from the stellar center, $c$ the speed of light.
Expanding the right-hand side with respect to
$\vert \mbox{\boldmath$\Omega$} \times \mbox{\boldmath$r$} \vert / c$,
we obtain 
\begin{equation}
  \rhoGJ 
  = -\frac{\Omega B_\zeta}{2\pi c}
           \left[ 1+O(\mbox{\boldmath$\Omega$} \times 
                      \mbox{\boldmath$r$}/c)^2
           \right],
  \label{eq:def_rhoGJ_1}
\end{equation}
where $B_\zeta \equiv \mbox{\boldmath$B$} \cdot
                      \mbox{\boldmath$\Omega$}/\Omega$.
For a Newtonian dipole magnetic field,
the null surface, where $B_\zeta$ and hence $\rhoGJ$ vanishes,
is located on a constant colatitude in the outer magnetosphere
on the poloidal plane
(heavy dashed line in fig.~\ref{fig:sideview}).

If the real charge density, $\rho_{\rm e}$,
deviates from $\rhoGJ$ in any region,
an electric field is exerted along $\mbox{\boldmath$B$}$.
If the potential drop is sufficient,
migratory electrons and/or positrons will be accelerated to radiate
$\gamma$-rays via curvature and/or inverse-Compton (IC) processes.
Some of such $\gamma$-rays collide with the soft photons 
illuminating the outer part of the magnetosphere
and materialize as pairs in the gap.
The replenished charges partially screen the original 
acceleration field,
$\Ell \equiv \mbox{\boldmath$E$}\cdot
             \mbox{\boldmath$B$}_{\rm p}/B_{\rm p}$,
where $\mbox{\boldmath$B$}_{\rm p}$ is the magnetic field
projected on the poloidal plane, 
and $B_{\rm p} \equiv \vert\mbox{\boldmath$B$}_{\rm p}\vert$.
If the created particles pile up at the boundaries of the potential gap,
they will quench the gap eventually.
Nevertheless, if the created particles continue to migrate
outside of the gap as a part of the global flows of charged particles,
a steady charge-deficient region could be maintained.
This is the basic idea of a particle acceleration zone 
in a pulsar magnetosphere.

In the CHR picture, the mainstream of the outer-gap model, 
the gap is hypothesized to be geometrically thin in the transfield
direction on the poloidal plane
in the sense $D_\perp \ll W$,
where $D_\perp$ represents the typical {\it transfield} 
thickness of the gap,
while $W$ does the width {\it along} the magnetic field lines.
In this limit, the acceleration electric field is partially screened 
by the zero-potential walls separated with a small distance $D_\perp$;
as a result, the gap, which is assumed to be vacuum,
extends from the null surface to (the vicinity of) 
the light cylinder (fig.~\ref{fig:sideview}). 

If $B_\zeta>0$ holds in the starward side of the null surface,
a positive acceleration field arises in the gap.
The light cylinder is defined as the surface
where the azimuthal velocity of a plasma would coincide with $c$
if it corotated with the magnetosphere.
Its distance, $\rlc \equiv c / \Omega$, 
from the rotational axis is called the \lq light cylinder radius'.
Particles are not allowed to migrate inwards beyond this surface
because of the causality in special relativity.

It should be noted that the null surface (where $B_\zeta$ vanishes)
is not a special place
for the gap electrodynamics in the sense that the plasmas are not
completely charge-separated in general and that the particles
freely pass through this surface inwards and outwards.
Therefore, the gap inner boundary is located near to the null surface,
not because a particle injection is impossible across this surface
(as previously discussed),
but because the gap is vacuum and transversely thin.

Then what happens in the CHR picture if the gap becomes 
no longer vacuum?
To consider this problem rigorously, we have to examine
the Poisson equation for the electrostatic potential. 
In fact, as will be explicitly demonstrated in the next section,
the original vacuum solution obtained in the pioneering work by
CHR cannot be applied to a non-vacuum CHR picture.
We are, therefore,  motivated by the need to solve
self-consistently the Poisson equation together with the
Boltzmann equations for particles and $\gamma$-rays.
Although the ultimate goal is to solve three-dimensional issues,
a good place to start is to examine one-dimensional problems.
In this context,
Hirotani and Shibata (1999a,~b,~c; hereafter Papers~I,~II,~III) 
and Hirotani (2000b, Paper~VI)
first solved the Boltzmann equations
together with the Maxwell equations
one-dimensionally along the field lines,
extending the idea originally developed for 
black-hole magnetospheres by Beskin et al. (1992).
Subsequently, Hirotani (2000a, Paper~IV; 2001, Paper~V)
considered a \lq gap closure condition' (eq.~[\ref{eq:closure}])
to constrain the gap width and estimated the $\gamma$-ray fluxes
for individual pulsars.

There is one important finding in this second picture:
The gap position shifts if there is a particle injection 
across either of the boundaries 
(Hirotani \& Shibata 2001, 2002a,b; hereafter Papers~VII, VIII, IX).
For example, 
when the injection rate across the outer (or inner) boundary
becomes comparable to the typical Goldreich-Julian value, 
the gap is located close to the neutron star surface 
(or to the light cylinder).
In other words, an outer gap is not quenched even 
when the injection rate of a completely charge-separated
plasma across the boundaries approaches the 
typical Goldreich-Julian value. 
Thus, an outer gap can coexist with a polar-cap accelerator;
this forms a striking contrast to the first, CHR picture.
It is also found in the second picture 
that an outer gap is quenched if the 
{\it created} particle density within the gap
exceeds some fractions of the Goldreich-Julian value.
That is, the {\it discharge} of created pairs is essential to screen
the acceleration field.

More recently, Hirotani, Harding, and Shibata (2003, Paper~X)
demonstrated that the particle energy
distribution cannot be represented 
either by a power law or by the mono-energetic approximation,
by solving explicitly the energy dependence of
particle distribution functions,
together with $\Ell$ and the $\gamma$-ray distribution functions.
They further showed that a soft power-law spectrum
is generally formed in 100~MeV-3~GeV energies
as a result of the superposition of the 
curvature spectra emitted by particles 
migrating at different positions.

In the present article,
we sum up the main points that have been made in the second picture
(from Papers~I to X),
which assumes that the gap is 
geometrically thick in the transfield direction
in the sense $D_\perp>0.3W$.

In the next section, 
we demonstrate that the charge distribution in 
the non-vacuum CHR model does not satisfy the Maxwell equation. 
We then analytically constrain the gap position in \S~\ref{sec:position},
and explicitly formulate the basic equations
in \S~\ref{sec:Vlasov} for quantitative analysis.
We apply the theory to individual pulsars in \S~\ref{sec:app}.
In the final section, 
we discuss the stability of the gap, 
evolution of the $\gamma$-ray luminosity vs. spin-down one,
and the unification of the CHR and the second pictures in outer-gap models,
as well as the unification of the outer-gap and the polar-cap models.


\section{Difficulties in Previous Outer-gap Models}
\label{sec:diffic}
To elucidate the electrodynamic difficulties in previous outer-gap
models, we have to examine the Poisson equation for the 
electrostatic potential.
In the inertial frame, the Poisson equation becomes
\begin{equation}
  \nabla \cdot \left[
    - \frac{1}{c}(\mbox{\boldmath$\Omega$}\times\mbox{\boldmath$r$})
      \times \mbox{\boldmath$B$} 
    - \nabla \Psi \right]
  = 4\pi \rho_{\rm e}(\mbox{\boldmath$r$}),
  \label{eq:Poisson_3D}
\end{equation}
where the left-hand side is the divergence of the electric field.
The real charge density, $\rho_{\rm e}$, 
is given by $\rho_{\rm e}= \rho_+ +\rho_-$, 
where $\rho_+$ and $\rho_-$ represent
the positronic and electronic charge densities, respectively.
The non-corotational potential $\Psi$ is related to the
usual scalar and vector potential ($A_0$,$\mbox{\boldmath$A$}$) as
$\Psi = A_0-(\Omega\varpi/c) 
            \mbox{\boldmath$e$}_\phi \cdot \mbox{\boldmath$A$}$,
where $\varpi$ designates the distance from the rotation axis,
$\mbox{\boldmath$e$}_\phi$ the azimuthal unit vector.
Noting that 
$ \mbox{\boldmath$E$}_\perp \equiv
   -(\mbox{\boldmath$\Omega$}\times\mbox{\boldmath$r$})
     \times \mbox{\boldmath$B$}/c$
represents the electric field perpendicular to the magnetic field,
we find that the term
$-\nabla^2 \Psi$ is important for particle acceleration 
in equation~(\ref{eq:Poisson_3D}).

Since the azimuthal dimension is supposed to be large
compared with $D_\perp$ in conventional outer-gap models, 
equation~(\ref{eq:Poisson_3D}) is reduced to the following 
two-dimensional form on the poloidal plane
\begin{equation} 
  -\frac{\partial^2 \Psi}{\partial s^2}
  -\frac{\partial^2 \Psi}{\partial z^2}
  = 4 \pi [\rho_{\rm e}(s,z)-\rhoGJ(s,z)],
  \label{eq:Poisson_2D}
\end{equation}
where equation~(\ref{eq:def_rhoGJ_0}) is used; 
$s$ and $z$ refer to the coordinates parallel and 
perpendicular, respectively, to the poloidal magnetic field. 
The star surface corresponds to $s=0$;
$s$ increases outwardly along the field lines.
The last-open field line corresponds to $z=0$; 
$z$ increases towards the magnetic axis (in the same hemisphere).
A dipole magnetic field has a single-signed curvature
near the last open field line (i.e., at $z \ll \rlc$)
except close to the light cylinder.
Thus, $\gamma$-rays propagate into the higher altitudes
(i.e., large $z$ regions); as a result,
in the CHR picture, the particle number density
$(\rho_+ -\rho_-)/e$ grows exponentially in the $z$ direction,
where $e$ designates the magnitude of the charge on an electron.
Because of this exponential growth of the particle number density,
it has been considered that most of the $\gamma$-rays are emitted
from the higher altitudes.

To explain the observed $\gamma$-ray luminosities with a small $D_\perp$,
one should assume that the created current density 
becomes the typical Goldreich-Julian value in the higher altitudes. 
That is, the conserved current density 
per magnetic flux tube should satisfy
\begin{equation}
  \frac{c\rho_+}{B} +\frac{-c\rho_-}{B}
  \sim \frac{\Omega}{2\pi}
  \label{eq:consv_0}
\end{equation}
in the order of magnitude, where $B \equiv \vert\mbox{\boldmath$B$}\vert$.
However, such a copious pair production will screen the 
local acceleration field, $\Ell = -\partial\Psi/\partial s$, 
as the Poisson equation~(\ref{eq:Poisson_2D}) indicates.

This screening effect is particularly important near to the
inner boundary.
Without loss of any generality, 
we can assume that $\Ell$ is positive.
In this case, because of the discharge, 
only electrons exist at the inner boundary. 
(We may notice here that external particle injections are not 
considered in the CHR picture.)
Thus, we obtain
$\rho_{\rm e}/B = \rho_-/B \sim -\Omega/(2\pi c)$
in the order of magnitude.
In the vicinity of the inner boundary,
we can Fourier-analyze equation~(\ref{eq:Poisson_2D})
in $z$ direction to find out that the $-\partial^2\Psi/\partial z^2$ term
contributes only to reduce 
$\partial\Ell/\partial s=-\partial^2\Psi/\partial s^2$.
Thus, a positive $-\rhoGJ$ must cancel the negative 
$\rho_{\rm e}$ to make the right-hand side be positive.
That is, at the inner boundary, 
\begin{equation}
  -\frac{\rhoGJ}{B}
  = \frac{\Omega}{2\pi c} \frac{B_\zeta}{B}
  > \frac{\vert \rho_{\rm e} \vert}{B}
  \sim \frac{\Omega}{2\pi c}
  \label{eq:inBDcond}
\end{equation}
must be satisfied,
so that the acceleration field may not change sign in the gap.
It follows that the polar cap, where $B_\zeta \sim B$ holds,
is the only place for the inner boundary of the \lq outer' gap 
to be located,
if the created particle number density in the gap is comparable to the
typical Goldreich-Julian value (eq.~[\ref{eq:consv_0}]). 
Such a non-vacuum gap must extend from the polar cap
(not from the null surface where $\rhoGJ$ vanishes)
to the light cylinder.
We can therefore conclude that 
the original vacuum solution obtained by
CHR cannot be applied to a non-vacuum CHR picture 
when there is a sufficient pair production that is needed 
to explain the observed $\gamma$-ray luminosity.

To construct a self-consistent model,
we have to solve equation~(\ref{eq:Poisson_2D})
together with the
Boltzmann equations for particles and $\gamma$-rays. 
In what follows, we consider this issue. 

\section{Analytic Examination of the Gap Position}
\label{sec:position}
Before turning to a closer examination of the particle
{\it Boltzmann} equations (after \S~\ref{sec:boltz_part}), 
it is helpful to describe the gap position and width
with the aid of particles {\it continuity} equations.
In this section, we analytically investigate this issue.

\subsection{Particle Continuity Equations}
\label{sec:cont}
At time $t$, position $\mbox{\boldmath$x$}$, 
and momentum $\mbox{\boldmath$p$}$, 
the distribution function $N$ of particles
obeys the following Boltzmann equation,
\begin{equation}
  \frac{\partial{N}}{\partial t}
    + \mbox{\boldmath$v$} \cdot \mbox{\boldmath$\nabla$} N
    + \left( e\mbox{\boldmath$E$}
             +\frac{\mbox{\boldmath$v$}}{c}
              \times\mbox{\boldmath$B$}
      \right) \cdot 
      \frac{\partial N}{\partial \mbox{\boldmath$p$}}
  = S(t,\mbox{\boldmath$x$},\mbox{\boldmath$p$}),
  \label{eq:boltz_1}
\end{equation}
where $\mbox{\boldmath$v$} \equiv $\mbox{\boldmath$p$}$/(m_{\rm e}\Gamma)$;
$m_{\rm e}$ refers to the rest mass of an electron,
and $\Gamma \equiv 1/\sqrt{1-(\vert \mbox{\boldmath$v$} \vert /c)^2}$
the Lorentz factor.
In a pulsar magnetosphere,
the collision term $S$ consists of the terms 
representing the particle appearing and disappearing 
rates at $\mbox{\boldmath$x$}$
and $\mbox{\boldmath$p$}$ per unit time per unit phase-space volume
due to pair production, pair annihilation, 
IC scatterings, and the synchro-curvature process.
Integrating equation~(\ref{eq:boltz_1}) over the momentum space,
and assuming that $N$ vanishes rapidly enough at
$p_i \rightarrow \pm \infty$ ($i=1,2,3$),
we obtain
\begin{equation}
  \frac{\partial \tilde{N}}{\partial t}
  + \mbox{\boldmath$\nabla$} \cdot 
    \left( \langle{\mbox{\boldmath$v$}}\rangle \tilde{N} \right)
  = \tilde{S}(t,{\mbox{\boldmath$x$}},{\mbox{\boldmath$p$}}),
  \label{eq:cont_1}
\end{equation}
where the particle number density $\tilde{N}$ and 
the averaged particle velocity 
$\langle{\mbox{\boldmath$v$}}\rangle$ are defined by
\begin{equation}
  \tilde{N}(t,{\mbox{\boldmath$x$}}) 
  \equiv \int_{-\infty}^\infty 
         N(t,{\mbox{\boldmath$x$}},{\mbox{\boldmath$p$}}) 
         d^3\mbox{\boldmath$p$},
  \quad
  \langle{\mbox{\boldmath$v$}}\rangle \equiv 
    \frac{\displaystyle{\int_{-\infty}^\infty 
                        \mbox{\boldmath$v$} N   d^3\mbox{\boldmath$p$}}}
         {\displaystyle{\int_{-\infty}^\infty N d^3\mbox{\boldmath$p$}}}.
  \label{eq:def_ave}
\end{equation}
Since the IC scatterings and the synchro-curvature
process conserve the particle number,
\begin{equation}
  \tilde{S}(t,{\mbox{\boldmath$x$}})
  \equiv \int_{-\infty}^\infty 
         S(t,{\mbox{\boldmath$x$}},{\mbox{\boldmath$p$}})
         d^3\mbox{\boldmath$p$}
  \label{eq:def_S}
\end{equation}
consists of pair production and annihilation terms.
For a typical pulsar magnetosphere,
the annihilation is negligibly small compared with the production.
Therefore, we obtain
\begin{equation}
  \tilde{S}(t,{\mbox{\boldmath$x$}}) =
  \frac{1}{c} \int_{0}^\infty dE_\gamma \, 
    \left[ \eta_{\rm p}(E_\gamma,\mu_+) G_+
          +\eta_{\rm p}(E_\gamma,\mu_-) G_- \right],
\end{equation}
where
$G_+(t,{\mbox{\boldmath$x$}},E_\gamma)$ and 
$G_-(t,{\mbox{\boldmath$x$}},E_\gamma)$ 
designate the distribution functions of
outwardly and inwardly propagating $\gamma$-ray photons,
respectively, having energy $E_\gamma$.
The pair-production redistribution functions are defined by
\begin{equation}
  \eta_{\rm p}(t,\mbox{\boldmath$x$};E_\gamma,\mu)
  = c
    \int_{-1}^{1} d\mu (1-\mu) 
     \int_{E_{\rm th}}^\infty dE_{\rm x}
     \frac{dN_{\rm x}}{dE_{\rm x}d\mu} 
     \sgP,
  \label{eq:def_etap_0}
\end{equation}
where $\sgP(E_\gamma,E_{\rm x},\mu)$ 
represents the pair-production cross section, and
\begin{equation}
  E_{\rm th} \equiv \frac{2}{1-\mu}\frac{(m_{\rm e}c^2)^2}{E_\gamma};
  \label{eq:def_Eth}
\end{equation}
$\cos^{-1}\mu_+$ (or $\cos^{-1}\mu_-$) is
the collision angle between the X-rays
and the outwardly (or inwardly) propagating $\gamma$-rays.

Since the drift motion due to the gradient and the curvature of 
$\mbox{\boldmath$B$}$ is negligible for typical outer-gap parameters,
we can decouple $\langle{\mbox{\boldmath$v$}}\rangle$ as
\begin{equation}
  \langle{\mbox{\boldmath$v$}}\rangle
  = \varpi \Omega_{\rm p} \mbox{\boldmath$e$}_\phi
   +c \cos\Phi \frac{\mbox{\boldmath$B$}_{\rm p}}{B_{\rm p}},
  \label{eq:decouple}
\end{equation}
where $\Omega_{\rm p}$ designates the angular velocity of
particles due to 
$\mbox{\boldmath$E$}_\perp \times \mbox{\boldmath$B$}_{\rm p}$ drift,
and $\Phi$ the projection angle of the 
particle three-dimensional motion onto the poloidal plane.
The drift angular velocity $\Omega_{\rm p}$ coincides $\Omega$
provided that $B_\phi=0$ and
$\vert\Ell\vert \ll \vert E_\perp \vert = B_{\rm p}\varpi/\rlc$ hold.
Imposing a stationary condition 
\begin{equation}
  \frac{\partial}{\partial t} 
  + \Omega_{\rm p} \frac{\partial}{\partial \phi} = 0,
  \label{eq:stationary}
\end{equation}
assuming 
$B_\phi \ll B_{\rm p} \equiv \vert \mbox{\boldmath$B$}_{\rm p} \vert$,
and utilizing $\mbox{\boldmath$\nabla$}\cdot\mbox{\boldmath$B$}=0$,
we obtain the following continuity equations for particles
from equation~(\ref{eq:cont_1})
\begin{equation}
  \pm B_{\rm p} \frac{\partial}{\partial s}
        \left( \frac{\tilde{N}_\pm}{B_{\rm p}} \right)
  = \frac{1}{\lambda_{\rm p} \vert\cos\Phi\vert} 
    \int_{0}^\infty dE_\gamma \left(G_+ +G_-\right),
  \label{eq:cont-eq}
\end{equation}
where $B_{\rm p} \partial/\partial s 
 =\mbox{\boldmath$B$}_{\rm p} \cdot \mbox{\boldmath$\nabla$}$.
Throughout this paper, we assume $\Ell>0$,
which does not lose any generality.
In this case, for positrons (or electrons), 
we have $\tilde{N}=\tilde{N}_+= \rho_+/e$ and 
$\cos\Phi=+\vert\cos\Phi\vert$
(or $\tilde{N}=\tilde{N}_-= -\rho_-/e$ and
    $\cos\Phi=-\vert\cos\Phi\vert$).
The pair-production mean free path $\lambda_{\rm p}(s)$ is defined by
\begin{equation}
  \frac{1}{\lambda_{\rm p}}
  \equiv 
  \frac{\displaystyle \int_0^\infty
                 \left[ \eta_{\rm p}(s,E_\gamma,\mu_+) G_+ 
                       +\eta_{\rm p}(s,E_\gamma,\mu_-) G_-
                 \right] dE_\gamma}
      {c\displaystyle \int_0^\infty (G_+(s,E_\gamma) +G_-(s,E_\gamma)) 
                      dE_\gamma}.
  \label{eq:def_mfp}
\end{equation}
Since the number of created positrons is always equal to that of
created electrons, the right-hand side of equation~(\ref{eq:cont-eq})
is common for $\tilde{N}_+$ and $\tilde{N}_-$. 
To examine the gap position, we must combine
equation~(\ref{eq:cont-eq}) with
the $\gamma$-ray Boltzmann equations.


\subsection{Elimination of Gamma-ray Distribution Functions}
\label{sec:eliminate}
In general, the $\gamma$-ray distribution function $G_\pm$ at momentum 
$\mbox{\boldmath$k$}$ obey the following Boltzmann equations
\begin{equation}
  \frac{\partial G_\pm}{\partial t} 
  + c\frac{\mbox{\boldmath$k$}}{\vert\mbox{\boldmath$k$}\vert}\cdot
    \nabla G_\pm(t,\mbox{\boldmath$x$},\mbox{\boldmath$k$}) 
  = S_\gamma(t,\mbox{\boldmath$x$},\mbox{\boldmath$k$}),
  \label{eq:Boltz_gam_0}
\end{equation}
where $S_\gamma$ represents the collision terms.
Unlike the charged particles,
$\gamma$-rays do not propagate along the magnetic field line at each point,
because they preserve the directional information where they were emitted.
However in this paper, we assume for simplicity
that they propagate along the local magnetic field lines
and put
\begin{equation}
  \frac{\mbox{\boldmath$k$}}{\vert\mbox{\boldmath$k$}\vert}
  = \cos\Phi_\gamma \frac{\mbox{\boldmath$B$}_{\rm p}}{B_{\rm p}}
   +\sin\Phi_\gamma \mbox{\boldmath$e$}_\phi.
  \label{eq:photon_mom}
\end{equation}
This assumption is justified 
if the magnetic field lines are nearly straight in the gap
(i.e., if $W \ll \rlc$ holds).
If aberration due to rotation is negligible, we obtain
$\Phi_\gamma=0$ (or $\pi$) for outwardly (or inwardly) propagating
$\gamma$-rays.
For simplicity, we further assume that the following
stationary condition is satisfied
\begin{equation}
  \left[ \frac{\partial}{\partial t}
        +c\sin\Phi_\gamma \frac{1}{\varpi}
                          \frac{\partial}{\partial \phi}
  \right] G_\pm = 0
  \label{eq:stationary_2}
\end{equation}
Then, noting that the curvature process is the dominant process
in $S_\gamma$, 
compared with the IC process, the pair creation,
and the pair annihilation for typical pulsar parameters,
we find that $G_\pm(s,E_\gamma)$ obey 
\begin{equation}
  c\cos\Phi_\gamma B_{\rm p}
  \frac{\partial}{\partial s}
  \left(\frac{G_\pm}{B_{\rm p}}\right)
  = \int_1^\infty d\Gamma \eta_{\rm c}(s,E_\gamma,\Gamma)
                          N_\pm(s,\Gamma),
  \label{eq:Boltz_gam_1}
\end{equation}
where (e.g., Rybicki, Lightman 1979)
\begin{equation}
  \eta_{\rm c}(s,E_\gamma,\Gamma) 
  \equiv \frac{\sqrt{3}e^2 \Gamma}{h \rho_{\rm c}(s)}
         \frac1{E_\gamma} 
         F \left( \frac{E_\gamma}{E_{\rm c}} \right),
  \label{eq:def-etaC}
\end{equation}
\begin{equation}
  E_{\rm c}(s,\Gamma) 
  \equiv \frac3{4\pi} \frac{hc \Gamma^3}{\rho_{\rm c}},
  \label{eq:def_Ec}
\end{equation}
\begin{equation}
  F(x) \equiv x \int_x^\infty K_{\frac53} (t) dt ;
\end{equation}
$\rho_{\rm c}(s)$ is the curvature radius of the magnetic field lines
and $K_{5/3}$ is the modified Bessel function of $5/3$ order.
The effect of the broad spectrum of curvature $\gamma$-rays
is represented by the factor $F(E_\gamma/E_{\rm c})$
in equation (\ref{eq:def-etaC}).

Integrating equation~(\ref{eq:Boltz_gam_1}) over $E_\gamma$,
combining with equations~(\ref{eq:cont-eq}), 
and assuming $\partial_s(\lambda_{\rm p}\cos\Phi)=0$,
we obtain
\begin{equation}
  \pm \frac{d^2}{ds^2} \left( \frac{N_\pm}{\Bp} \right)
  = \frac{1}{\vert\cos\Phi\cos\Phi_\gamma\vert}
    \frac{1}{\lambda_{\rm p} c} \frac{N_+ -N_-}{\Bp}
    \int_0^\infty \eta_{\rm c} dE_\gamma,
  \label{eq:cont-eq3}
\end{equation}
where $0<\Phi_\gamma<\pi/2$ (or $\pi/2<\Phi_\gamma<\pi$)
is applied to outwardly (or inwardly) propagating $\gamma$-rays.

\subsection{Real Charge Density in the Gap}
\label{sec:rhoe}
One combination of the two independent equations~(\ref{eq:cont-eq3})
yields the current conservation law; 
that is, the total current density per magnetic flux tube,
\begin{equation} 
  j_{\rm tot}
  = \frac{2\pi ce}{\Omega}
    \frac{\tilde{N}_+(s) +\tilde{N}_-(s)}{\Bp(s)}
  \label{eq:consv_1}
\end{equation}
is conserved along the field lines.
(Note that it can be derived directly from eq.~[\ref{eq:cont-eq}].)
Another combination gives
\begin{equation} 
  \frac{d^2}{ds^2} \left(\frac{\tilde{N}_+ -\tilde{N}_-}{\Bp}\right)
  = \frac{2}{W}\cdot
    \frac{N_\gamma}{\lambda_{\rm p}\vert\cos\Phi_\gamma\vert} 
    \frac{\tilde{N}_+ -\tilde{N}_-}{\Bp},
  \label{eq:master_eq_1}
\end{equation}
where 
\begin{equation}
  N_\gamma 
  \equiv 
  \frac{W}{c\vert\cos\Phi\vert} 
  \int_0^\infty \eta_{\rm c}(s,\Gamma,E_\gamma) dE_\gamma
  \label{eq:Ngamma}
\end{equation}
refers to the expectation value of the number of $\gamma$-rays
emitted by a single particle that runs the gap width, $W$.
Lorentz factor appearing in $\eta_{\rm c}$ should be
evaluated at each position $s$.

Exactly speaking, $\lambda_{\rm p}$ depends on $G_+$ and $G_-$;
thus, the $\gamma$-ray distribution functions are not
eliminated in equation~(\ref{eq:cont-eq3}).
Nevertheless, for analytic (and qualitative) discussion of 
the gap position,
we may ignore such details and adopt equation~(\ref{eq:master_eq_1}).

A typical $\gamma$-ray propagates the length 
$W/(2\vert\cos\Phi_\gamma\vert)$ 
within the gap that is transversely thick.
Thus, so that a stationary pair-production cascade may be maintained,
the optical depth, $W/(2\vert\cos\Phi_\gamma\vert\lambda_{\rm p})$, 
must equal the 
expectation value for a $\gamma$-ray to materialize with the gap,
$N_\gamma^{-1}$.
We thus obtain the following condition:
$W/2=\vert\cos\Phi_\gamma\vert\lambda_{\rm p}/N_\gamma$.
This relation holds for a self-sustaining gap in which
all the particles are supplied by the pair production.
If there is an external particle injection, 
the injected particles also contribute for the $\gamma$-ray emission.
As a result, a stationary gap can be maintain with a 
smaller width compared to the case of no particle injection.
Taking account of such injected particles,
we can constrain the half gap width as
\begin{equation}
  \frac{W}{2}
  = \frac{\lambda_{\rm p}\vert\cos\Phi_\gamma\vert}{N_\gamma}\cdot
    \frac{j_{\rm gap}}{j_{\rm tot}},
  \label{eq:closure}
\end{equation}
where $j_{\rm gap}$ and $j_{\rm tot}$ refer to the created and total
current densities per unit magnetic flux tube.
Equation~(\ref{eq:closure}) is automatically satisfied 
if we solve the set of Maxwell and stationary Boltzmann equations.
Here, $j_{\rm gap}$ is related with the particle injection rate across the
boundaries as follows:
\begin{eqnarray}
  \frac{\Omega}{2\pi ce} j_{\rm gap} 
  &=& \frac{\tilde{N}_+(s^{\rm out})}{\Bp(s^{\rm out})}
     -\frac{\tilde{N}_+(s^{\rm in })}{\Bp(s^{\rm in })}
  \nonumber\\
  &=& \frac{\tilde{N}_-(s^{\rm in })}{\Bp(s^{\rm in })}
     -\frac{\tilde{N}_-(s^{\rm out})}{\Bp(s^{\rm out})},
  \label{eq:def_jgap}
\end{eqnarray}
where $s^{\rm in}$ and $s^{\rm out}$ designate the position of the
inner and the outer boundaries, respectively.
That is, $W= s^{\rm out} -s^{\rm in}$.

With the aid of identity~(\ref{eq:closure}),
we can rewrite equation~(\ref{eq:master_eq_1}) into the form
\begin{equation} 
  \frac{d^2}{ds^2} \left(\frac{\tilde{N}_+ -\tilde{N}_-}{\Bp}\right)
  = \frac{j_{\rm gap}}{j_{\rm tot}}
    \frac{4}{W^2} \frac{\tilde{N}_+ -\tilde{N}_-}{\Bp}.
  \label{eq:master_eq}
\end{equation}
To solve this differential equation,
we impose the following two boundary conditions:
\begin{equation}
  ce \frac{\tilde{N}_+(s^{\rm in})}{\Bp(s^{\rm in})}
  = \frac{\Omega}{2\pi} j^{\rm in},
  \label{eq:BD_extra_1}
\end{equation}
\begin{equation}
  ce \frac{\tilde{N}_-(s^{\rm out})}{\Bp(s^{\rm out})}
  = \frac{\Omega}{2\pi} j^{\rm out}.
  \label{eq:BD_extra_2}
\end{equation}
Then, equation~(\ref{eq:def_jgap}), 
(\ref{eq:BD_extra_1}), and (\ref{eq:BD_extra_2}) give
\begin{equation}
  \frac{\tilde{N}_+ -\tilde{N}_-}{\Bp}
  = -\frac{\Omega}{2\pi ce}(j_{\rm gap}-j^{\rm in}+j^{\rm out})
  \label{eq:BD_01}
\end{equation}
at $s= s^{\rm in}$, and 
\begin{equation}
  \frac{\tilde{N}_+ -\tilde{N}_-}{\Bp}
  = \frac{\Omega}{2\pi ce}(j_{\rm gap}+j^{\rm in}-j^{\rm out})
  \label{eq:BD_02}
\end{equation}
at $s= s^{\rm out}$.
Under boundary conditions~(\ref{eq:BD_01}) and (\ref{eq:BD_02}),
equation~(\ref{eq:master_eq}) is solved as
\begin{eqnarray}
  \frac{\tilde{N}_+ -\tilde{N}_-}{\Bp}
  &=& \frac{\Omega}{2\pi ce}
      \left[ j_{\rm gap} 
             \frac{\sinh\left(\sqrt{\displaystyle\frac{j_{\rm gap}}
                                                      {j_{\rm tot}}
                                   }
                               \displaystyle\frac{s-s_{\rm cnt}}
                                                {W/2}
                        \right)
                  }
                  {\sinh\left(\sqrt{\displaystyle\frac{j_{\rm gap}}
                                                      {j_{\rm tot}}
                                   }
                        \right)
                  }
      \right.
  \nonumber\\
  & &
  \hspace*{-2.0 truecm}
  + \left. (j^{\rm in}-j^{\rm out})
             \frac{\cosh\left(\sqrt{\displaystyle\frac{j_{\rm gap}}
                                                      {j_{\rm tot}}
                                   }
                               \displaystyle\frac{s-s_{\rm cnt}}
                                                {W/2}
                        \right)
                  }
                  {\cosh\left(\sqrt{\displaystyle\frac{j_{\rm gap}}
                                                      {j_{\rm tot}}
                                   }
                        \right)
                  }
      \right],
  \label{eq:charge_density}
\end{eqnarray}
where the gap center position is defined by
\begin{equation}
  s_{\rm cnt} \equiv \frac{s^{\rm in}+s^{\rm out}}{2}.
  \label{eq:def_scnt}
\end{equation}

Note that $e(\tilde{N}_+ -\tilde{N}_-)=\rho_{\rm e}$
represents the real charge density,
which appears in the Poisson equation.
Thus, substituting equation~(\ref{eq:charge_density})
into (\ref{eq:Poisson_2D}), we obtain
\begin{eqnarray}
  -\nabla^2 \Psi 
    &=& \frac{2\Bp\Omega}{c}
        \left[ j_{\rm gap} 
                  f_{\rm odd} \left( \frac{s-s_{\rm cnt}}{W/2} \right)
        \right.
    \nonumber\\
    & & \hspace*{-1.0 truecm}
        \left.
              +(j^{\rm in}-j^{\rm out})
                  f_{\rm even}\left( \frac{s-s_{\rm cnt}}{W/2} \right)
              +\frac{B_\zeta}{B}
        \right],
  \label{eq:Poisson_1}
\end{eqnarray}
where 
\begin{equation}
  f_{\rm odd}(x) 
  \equiv \frac{\sinh\left( x \sqrt{j_{\rm gap}/j_{\rm tot}} \right)}
              {\sinh\left(   \sqrt{j_{\rm gap}/j_{\rm tot}} \right)},
  \label{eq:Poisson_1a}
\end{equation}
\begin{equation}
  f_{\rm even}(x) 
  \equiv \frac{\cosh\left( x \sqrt{j_{\rm gap}/j_{\rm tot}} \right)}
              {\cosh\left(   \sqrt{j_{\rm gap}/j_{\rm tot}} \right)}.
  \label{eq:Poisson_1b}
\end{equation}

At the inner boundary, $s=s^{\rm in}$,
$(s-s_{\rm cnt})/(W/2)=-1$ holds;
therefore, we obtain
\begin{equation}
  -\nabla^2 \Psi
  = \frac{2\Bp\Omega}{c} 
    \left( -j_{\rm gap}+j^{\rm in}-j^{\rm out}+\frac{B_\zeta}{\Bp}
    \right).
  \label{eq:poisson_inner}
\end{equation}
In the CHR picture, it is assumed that there is no particle
injection across either of the boundaries 
(i.e., $j^{\rm in}=j^{\rm out}=0$)
and that the current density associated with the created particles
becomes of the order of the typical Goldreich-Julian value
(i.e., $j_{\rm gap} \sim 1$).
It follows that the inner boundary of an \lq outer gap'
should be located close to the star,
where $B_\zeta \sim B_{\rm p}$ holds;
this conclusion is consistent with what was obtained 
in \S~\ref{sec:diffic}.

\subsection{Gap Position vs. Particle Injection}
\label{sec:null}

To examine the Poisson equation~(\ref{eq:Poisson_1}) analytically,
we assume that the transfield thickness of the gap is 
greater than $W$ and replace $\nabla^2 \Psi$ with $d^2 \Psi/ds^2$.
Furthermore, we neglect the current created in the gap
and put $j_{\rm gap} \sim 0$.

First, consider the case when particles are injected 
across neither of the boundaries(i.e., $j^{\rm in}=j^{\rm out}=0$).
It follows that the
derivative of the $\Ell$ vanishes
at the null surface, where $B_\zeta$ vanishes.
We may notice that
$-d^2\Psi/ds^2=d\Ell/ds$ is positive at the inner part of the gap
and becomes negative at the outer part.
The acceleration field is screened out at the
boundaries by virtue of the spatial distribution of the
local Goldreich-Julian charge density, $\rhoGJ$.
Therefore, we can conclude that the gap is located 
(or centers) around the null surface,
if there is no particle injection from outside.
This conclusion can be easily generalized to the case
$j^{\rm in}=j^{\rm out}\ne 0$.

Secondly, consider the case when particles are injected 
across the inner boundary at $s=s^{\rm in}$
(or in general, when $j^{\rm in}-j^{\rm out}>0$ holds).
Since the function $f_{\rm even}$ is positive at arbitrary $s$,
the gap center is located at a place where $B_\zeta$ is negative,
that is, outside of the null surface. 
In particular, when $j^{\rm in}-j^{\rm out} \sim 1$ holds, 
$d\Ell/ds$ vanishes at the place where $B_\zeta \sim -B$ holds.
In a vacuum, static dipole field,
$B_\zeta \sim -B$ is realized along the last-open field line
near to the light cylinder.
Therefore, the gap should be located close to the light cylinder,
if the injected particle flux across the inner boundary
approaches the typical Goldreich-Julian value. 
We may notice here that $f_{\rm even}$ is less than unity,
because $\vert s-s_{\rm cnt} \vert$ does not exceed $W/2$.

Thirdly and finally, consider the case 
when $j^{\rm in}-j^{\rm out} \sim -1$ holds.
In this case, $d\Ell/ds$ vanishes at the place where $B_\zeta \sim B$.
Therefore, an \lq outer' gap should be located in the polar cap,
if a Goldreich-Julian particle flux is injected across the outer boundary. 

\section{The Set of Maxwell and Boltzmann Equations}
\label{sec:Vlasov}
To examine the gap electrodynamics more quantitatively,
we have to solve numerically
the Poisson equation for the electrostatic potential 
together with the Boltzmann equations for particles and $\gamma$-rays.
To this aim, we reduce these equations to a tractable forms
in this section.

\subsection{One-dimensional Poisson Equation}
\label{sec:Poisson_1D}
For simplicity, we assume that a gap is transversely thick
in the sense $D_\perp > W$ (or at least $D_\perp \sim W$).
In this case, the derivative with respect to $z$
in the left-hand side of the 
Poisson equation~(\ref{eq:Poisson_2D}) 
can be approximated with $-\Psi/D_\perp^2$.
Thus, we obtain the following one-dimensional expression 
of the Poisson equation~(\ref{eq:Poisson_2D}):
\begin{equation} 
  -\frac{\partial^2 \Psi}{\partial s^2}
  = -\frac{\Psi}{D_\perp^2}
    + 4 \pi \left[\rho_{\rm e}(s)+\frac{\Omega B_\zeta(s)}{2\pi c}
            \right],
  \label{eq:BASIC_1}
\end{equation}
where 
\begin{equation}
  \rho_{\rm e}(s) 
  = e \int_1^\infty d\Gamma 
      \left[ N_+(s,\Gamma) -N_-(s,\Gamma) \right].
  \label{eq:def_rhoe}
\end{equation}
The particle distribution functions $N_+$ and $N_-$ obey
the Boltzmann equations that will be described just below. 

\subsection{Particle Boltzmann Equations}
\label{sec:boltz_part}
In this section, we consider the Boltzmann equations~(\ref{eq:boltz_1}),
which is necessary to investigate energy distribution of particles.
It should be noted that 
quantum effects can be neglected in the outer magnetosphere,
because the magnetic field is much less than the critical value 
($4.41 \times 10^{13}$~G).
As a result, synchro-curvature radiation takes place continuously and 
can be regarded as an external force acting on a particle.
If we instead put the collision term associated with 
the synchro-curvature process in the right-hand side, 
the energy transfer in each collision would be too small
to be resolved by the energy grids.

In the same manner as we derived the stationary continuity
equations~(\ref{eq:cont-eq}),
we impose the stationary condition~(\ref{eq:stationary}).
Then, neglecting the pitch-angle dependence of the
particle distribution functions,
and approximating the collision term associated with the curvature
process as an external force acting on a particle,
we obtain (Appendix~A in Paper~X)
\begin{equation}
   \Bp\frac{\partial}{\partial s}\left(\frac{N_+}{\Bp}\right)
   +\left[ e\Ell 
          -\frac{\Pcv(s,\Gamma)}{c} \right]
    \frac{\partial N_+}{\partial\Gamma}
  =  S_+(s,\Gamma),
 \label{eq:BASIC_2a}
\end{equation}
\begin{equation}
  \Bp\frac{\partial}{\partial s}\left(\frac{N_-}{\Bp}\right)
   -\left[ e\Ell 
          -\frac{\Pcv(s,\Gamma)}{c} \right]
    \frac{\partial N_-}{\partial\Gamma}
  = -S_-(s,\Gamma),
 \label{eq:BASIC_2b}
\end{equation}
where the radiation-reaction force is given by 
\begin{equation}
  \frac{\Pcv(s,\Gamma)}{c}
  = \frac{2e^2 \Gamma^4}{3\rho_{\rm c}^2(s)};
  \label{def_Pcv}
\end{equation}
$\rho_{\rm c}$ is the curvature radius of the magnetic field line.

\subsubsection{Collision terms}
\label{sec:coll_terms}
We assume in this paper that $\gamma$-rays are either outwardly or
inwardly propagating along the local magnetic field lines.
We also assume that the soft photons are emitted from the 
neutron star and hence unidirectional at the gap.
Then the cosine of the collision angle $\mu$ has a unique value
$\mu_+$ or $\mu_-$, for outwardly or inwardly propagating
$\gamma$-rays, respectively;
$\mu_+$ and $\mu_-$ are determined by the
magnetic inclination $\inc$ at each position $s$ (eq.~[23] in Paper~VII).
Under these assumptions,
the source term can be expressed as
\begin{eqnarray}
  \lefteqn{ S_+(s,\Gamma) 
      = - \int_{E_\gamma<\Gamma} dE_\gamma \etaICg(E_\gamma,\Gamma,\mu_+) 
                                 N_+(s,\Gamma)}
  \nonumber \\  
  &+& \int_{\Gamma_i>\Gamma} d\Gamma_i \, 
           \etaICe (\Gamma_i, \Gamma, \mu_+) N_+(s,\Gamma_i)
     +Q_{\rm p}(s,\Gamma),
  \nonumber \\  
   \label{eq:src_1}
\end{eqnarray}
\begin{eqnarray}
  \lefteqn{ S_-(s,\Gamma) 
      = - \int_{E_\gamma<\Gamma} dE_\gamma \etaICg(E_\gamma,\Gamma,\mu_-) 
                                 N_-(s,\Gamma)}
  \nonumber \\  
  &+& \int_{\Gamma_i>\Gamma} d\Gamma_i \, 
           \etaICe (\Gamma_i, \Gamma, \mu_-) N_-(s,\Gamma_i)
     +Q_{\rm p}(s,\Gamma),
  \nonumber \\  
   \label{eq:src_2}
\end{eqnarray}
where the pair-production rate per unit volume per unit Lorentz factor
is defined as
\begin{equation}
  Q_{\rm p} \equiv 
  \int dE_\gamma 
      \left[ \frac{\partial \etaP(E_\gamma,\Gamma,\mu_+)}
                  {\partial \Gamma} G_+
            +\frac{\partial \etaP(E_\gamma,\Gamma,\mu_-)}
                  {\partial \Gamma} G_-
                \right].
  \label{eq:def_Qp}
\end{equation}
The positrons (or electrons) are supposed to
collide with the soft photons at the same angle as the
outwardly (or inwardly) propagating $\gamma$-rays;
therefore, the same collision angle $\cos^{-1}\mu_+$ (or $\cos^{-1}\mu_-$)
is used for both IC scatterings and pair production in 
equations (\ref{eq:src_1})--(\ref{eq:def_Qp}).
The collisions tend to be head-on (or tail-on) 
for inwardly (or outwardly) propagating $\gamma$-rays, 
as the gap approaches the star.
The pair-production redistribution function is given by
\begin{equation}
  \frac{\partial\etaP}{\partial\Gamma}(E_\gamma,\Gamma,\mu_\pm)
  = (1-\mu_\pm)
    \int_{E_{\rm th}}^\infty dE_{\rm s}
    \frac{dF_{\rm s}}{dE_{\rm s}}
    \frac{d\sgP}{d\Gamma},
  \label{eq:def_etaP_2}
\end{equation}
where the pair-production threshold energy is defined by
equation~(\ref{eq:def_Eth}).
The differential cross section 
$d\sigma_{\rm p}/d\Gamma$ is given in numerous textbooks in
quantum electrodynamics
(e.g., Akhiezer \& Berestetskii 1965; or eq.~[32] in Paper~X)

The IC redistribution function 
$\etaICg(E_\gamma,\Gamma,\mu)$ represents the probability
that a particle with Lorentz factor $\Gamma$ upscatters photons 
into energies between $E_\gamma$ and $E_\gamma+dE_\gamma$ 
per unit time when the collision angle is $\cos^{-1}\mu$.
On the other hand, $\etaICe(\Gamma_i,\Gamma,\mu)$ describes
the probability that a particle changes Lorentz factor from
$\Gamma_i$ to $\Gamma$ in a scattering.
Thus, energy conservation gives
\begin{equation}
  \etaICe(\Gamma_i,\Gamma_f,\mu) 
     = \etaICg[m_{\rm e}c^2(\Gamma_i-\Gamma_f),\Gamma_i,\mu]
  \label{eq:rel_etaIC}
\end{equation}

In general, $\etaICg$ is defined by 
the soft photon flux $dF_{\rm s}/d\Es$ and 
the Klein-Nishina cross section $\sigma_{\rm KN}$ as follows: 
\begin{eqnarray}
  \lefteqn{\etaICg(E_\gamma,\Gamma,\mu_\pm)
            = (1-\beta\mu_\pm)}
  \nonumber \\
  &\times&
      \int_0^\infty dE_{\rm s} \frac{dF_{\rm s}}{dE_{\rm s}}
       \int_{-1}^{1} d\Omega_\gamma^\ast 
            \frac{d\sigma_{\rm KN}^\ast}{d{E_\gamma}^\ast 
            d\Omega_\gamma^\ast}
       \frac{d{E_\gamma}^\ast}{dE_\gamma}
  \nonumber \\
  \label{eq:def_etaICg_1}
\end{eqnarray}
where $\beta \equiv \sqrt{1-1/\Gamma^2}$ is virtually unity, 
$\Omega_\gamma$ the solid angle of upscattered photon,
the asterisk denotes the quantities in the electron (or positron) 
rest frame.
In the rest frame of a particle,
a scattering always takes place well above the resonance energy.
Thus, the classical formula of the Klein-Nishina cross section
can be applied to the present problem.
The soft photon flux per unit 
photon energy $E_{\rm s}$ [$s^{-1}\mbox{cm}^{-2}\mbox{ergs}^{-1}$] 
is written as $dF_{\rm s}/dE_{\rm s}$.
To obtain $\etaICg$'s for individual pulsars,
we substitute the observed X-ray spectrum 
$dF_{\rm s}/dE_{\rm s}$ and execute integration over
$E_{\rm s}$ and $\Omega_\gamma^\ast$.
For further details of $\etaICg$, see Appendix~B in Paper~X.


\subsection{Gamma-ray Boltzmann Equations}
\label{sec:boltz_gamma}
Let us briefly comment on the $\gamma$-ray Boltzmann equations.
We recover $\gamma$-ray production due to IC scatterings 
and absorption due to pair production in the right-hand side of
equation~(\ref{eq:Boltz_gam_1}) to obtain
\begin{eqnarray}
  \lefteqn{\pm c\vert\cos\Phi_\gamma\vert
           \Bp \frac{\partial}{\partial s} 
                  \left( \frac{G_\pm}{\Bp}\right)
     = - \int d\Gamma \frac{\partial\etaP}{\partial\Gamma} 
         \cdot G_\pm(s,E_\gamma)}
  \nonumber \\
  &+& \int_1^\infty 
      d\Gamma \left[ \etaICg(E_\gamma,\Gamma,\mu_\pm)
                    +\eta_{\rm c}( E_\gamma,\Gamma)
              \right]
              N_\pm(s,\Gamma).
  \nonumber \\
  \label{eq:BASIC_3}
\end{eqnarray}   
We integrate both sides of equation~(\ref{eq:BASIC_3})
over $E_\gamma$ in appropriate energy bins to reduce them into 
ordinary differential equations.

In short, the set of Maxwell and Botzmann equations 
consist of equations~(\ref{eq:BASIC_1}),
(\ref{eq:BASIC_2a}), (\ref{eq:BASIC_2b}), and (\ref{eq:BASIC_3}).
The Poisson equation and the $\gamma$-ray Boltzmann equations
become ordinary differential equations,
which can be straightforwardly solved by a simple discretization.
On the other hand, the hyperbolic-type partial differential 
equations~(\ref{eq:BASIC_2a}) and (\ref{eq:BASIC_2b})
are solved by the Cubic Interpolated Propagation (CIP) scheme
(e.g., Yabe \& Aoki 1991, Yabe, Xiao, \& Utsumi 2001).

\subsection{Boundary Conditions}
\label{sec:BD}
In this section, we consider the boundary conditions 
to solve the set of Maxwell and Boltzmann equations.
Diving the $\gamma$-ray energies into $m$ bins,
and diving the Lorentz factors into $n$ bins,
we impose the following boundary conditions 
at the {\it inner} (starward) boundary ($s= s^{\rm in}$)
\begin{equation}
  \Ell(s^{\rm in})=0,
  \quad
  \psi(s^{\rm in}) = 0,
  \label{eq:BD-1}
\end{equation}
\begin{equation}
  \int_{b_i}^{b_{i+1}} G_+(s^{\rm in},E_\gamma) dE_\gamma =0  
  \quad \mbox{for} \quad i= 1, 2, \ldots m
  \label{eq:BD-3}
\end{equation}
and
\begin{equation}
  N_+(s^{\rm in},\Gamma_j) 
  = \frac{\Omega\Bp(s^{\rm in})}{2\pi ce} y(\Gamma_j)
  \quad \mbox{for} \quad j= 1, 2, \ldots n,
  \label{eq:BD-4}
\end{equation}
where $y$ is an appropriate function satisfying
$\int_1^\infty y(\Gamma)d\Gamma=j^{\rm in}$.
Moreover, current conservation law~(\ref{eq:consv_1}) gives
\begin{equation}
  \int_1^\infty N_-(s^{\rm in},\Gamma) d\Gamma
  = \frac{\Omega\Bp(s^{\rm in})}{2\pi ce} (j_{\rm tot}-j^{\rm in}).
  \label{BD-5}
\end{equation}

At the {\it outer} boundary ($s=s^{\rm out}$), we impose
\begin{equation}
  \Ell(s^{\rm out})=0,
  \label{eq:BD-6}
\end{equation}
\begin{equation}
  \int_{b_i}^{b_{i+1}} G_-(s^{\rm out},E_\gamma) dE_\gamma =0  
  \quad \mbox{for} \quad i= 1, 2, \ldots m
  \label{eq:BD-7}
\end{equation}
\begin{equation}
  N_-(s^{\rm out},\Gamma_j) 
  = \frac{\Omega\Bp(s^{\rm out})}{2\pi ce} y(\Gamma_j)
  \quad \mbox{for} \quad j= 1, 2, \ldots n.
  \label{eq:BD-8}
\end{equation}
The current density created in the gap per unit flux tube
can be expressed as
\begin{equation}
  j_{\rm gap}= j_{\rm tot} -j^{\rm in} -j^{\rm out}.
  \label{eq:Jgap}
\end{equation}
We adopt $j_{\rm gap}$, $j^{\rm in}$, and $j^{\rm out}$
as the free parameters.
We chose $y(\Gamma)$ so that the initial spectrum of the
injected particles may peak near the lowest energy bin
(e.g., solid curve in fig.~\ref{fig:PVela_75a});
then, the results little depend on the detailed form of $y(\Gamma)$.

We have totally $2m+2n+4$ boundary conditions 
(\ref{eq:BD-1})--(\ref{eq:BD-8})
for $2m+2n+2$ unknown functions
$\int_{b_i}^{b_{i+1}}G_\pm(s,E_\gamma)dE_\gamma$,
$N_\pm(s,\Gamma_j)$, $\Psi(s)$, and $\Ell(s)$.
Therefore, two extra boundary conditions must be compensated 
by making the positions of the boundaries 
$s^{\rm in}$ and $s^{\rm out}$ be free.
The two free boundaries appear because $\Ell=0$ is imposed at 
{\it both} the boundaries and because $j_{\rm gap}$ is externally imposed.
In other words, the gap boundaries 
($s^{\rm in}$ and $s^{\rm out}$) shift,
if $j^{\rm in}$ and/or $j^{\rm out}$ varies.
That is, the gap position, as well as its width $W$,
cannot be artificially hypothesized as in previous outer-gap models.
They should be self-consistently solved from the set of
Maxwell and Boltzmann equations.

\section{Application to Individual Pulsars}
\label{sec:app}
To solve the set of the Maxwell and Boltzmann equations,
we must specify the
X-ray field, $dF_{\rm s}/d\Es$, which is necessary to
compute the pair-production redistribution function
(eq.~[\ref{eq:def_etaP_2}]).
In this paper, we use the X-ray fluxes and spectra
observed for individual rotation-powered pulsars.
In \S~\ref{sec:app_1}, we summarize the observed properties
of the X-ray field.
Then we apply the theory to
the Vela pulsar in \S~\ref{sec:app_vela},
to PSR~B1706-44 in \S~\ref{sec:app_1706},
to the Geminga pulsar in \S~\ref{sec:app_Gemi},
and to PSR~B1055-52 in \S~\ref{sec:app_1055}.
We assume that the solid angle of the emitted $\gamma$-rays is 1~ster
throughout this paper.

\subsection{Input Soft Photon field}
\label{sec:app_1}
We consider the photons emitted from the
neutron star surface as the seed photons for
($\gamma$-$\gamma$) pair-production and IC scatterings.
That is, we do not consider power-law X-ray components,
because they are probably magnetospheric and beamed away from 
the accelerator.
We evaluate the IR photon field, which is needed to compute the
IC scattering rate, from the Rayleigh-Jeans tail of the surface
thermal component.
In table~1, we present the observed properties of the 
four $\gamma$-ray pulsars
exhibiting surface X-ray components,
in order of spin-down luminosity, $L_{\rm spin}$.

\begin{table*}
  \centering
    \begin{minipage}{160mm}
      \caption{Input thermal X-ray field}
      \begin{tabular}{@{}lccccrcrcll@{}}
        \hline
        \hline
        pulsar	
		& $\lg L_{\rm spin}$
		& distance
		& $\Omega$		& $\lg B_{\rm s}$
		& $kT_{\rm s}$		& $A_{\rm s}/A_\ast{}^\dagger$
		& $kT_{\rm h}$		& $A_{\rm h}/A_\ast{}^\dagger$
		& model
		& refs.				\\
        \	
		& ergs s${}^{-1}$
		& kpc
		& rad s${}^{-1}$	& G
		& eV			&
		& eV			& 
		& 
		&					\\
        \hline
        Vela	
		& 36.84		
		& 0.25
		& 70.4		& 12.53
		& 59		& 1.000
		& $\ldots$	& $\ldots$
		& hydrogen atm.
		& 1 				\\
        B1706-44
		& 36.53
		& 2.50
		& 61.3		& 12.49
		& 143		& 0.129
		& $\ldots$	& $\ldots$
 		& blackbody
		& 2					\\
        Geminga
		& 34.51		
		& 0.16
		& 26.5		& 12.21
		& 50		& 0.208
		& $\ldots$	& $\ldots$
 		& blackbody
		& 3, 4				\\
        B1055--52
		& 34.48		
		& 1.53
		& 31.9		& 12.03
		& 78		& 1.340
		& 740		& $10^{-5.32}$
 		& blackbody
		& 5					\\
        \hline
      \end{tabular}
      \begin{flushleft}
      ${}^\dagger$ 
        $A_\ast=4\pi(10\mbox{ km})^2$;
      \quad
    1: Pavlov et al. 2001;                  
    \quad
    2: Gotthelf, Halpern, Dodson 2002;      
    \quad
    3:  Halpern \& Wang 1997;                
    \quad
    4:  Becker \& Tr$\ddot{\rm u}$mper 1996; 
    \quad
    5: Mineo et al. 2002.           
    \quad
      \end{flushleft}
    \end{minipage}
\end{table*}

{\bf Vela} (J0835--4513)\ 
From Chandra observations in 0.25-8.0~keV,
the spectrum of this pulsar is turned out to consist of
two distinct component:
A soft, thermal component and a hard, power-law component.
As stated just above, we consider only the former component
as the X-ray field illuminating the outer gap.
This component can be modeled 
as a magnetic hydrogen atmosphere spectrum
with effective temperature $kT=0.68$MK
(Pavlov et al.~2001). 
Based on high-resolution Ca~{\small II} and Na~{\small I}
absorption-line spectra toward 68 OB stars in the direction
of the Vela supernova remnant,
Cha, Sembach, and Danks (1999) determined the distance to be
$250 \pm 30$~pc.\\
{\bf B1706--44} (J1710--4432)\ 
Gotthelf, Halpern, Dodson (2002) reported
a broad, single-peaked pulsed profile with pulsed fraction of $23 \%$,
using the High Resolution Camera on-board the Chandra X-ray observatory.
They fitted the spectroscopic data to find
(at least) two components: 
A blackbody of $kT=143$~eV with $A= 0.129 A_\ast$
and a power-law component with photon index of $-2.0$,
where $A_\ast=4\pi(10\mbox{ km})^2$.
We consider that the former component illuminates the gap efficiently
and neglect the latter one.
We adopt $d=2.5$~kpc as a compromise between the smaller 
dispersion-measure distance of $1.8$~kpc based on the
free electron model by Taylor and Cordes (1993)
and the larger H~I kinematic distance of $2.4$--$3.2$~kpc
derived by Koribalski et al.~(1995).\\
{\bf Geminga} (J0633+1746)\ 
The X-ray spectrum consists of two components:
the soft surface blackbody with $kT_{\rm s}=50$ eV and 
$A_{\rm s}= 0.21 A_* (d/0.16)^2$
and a hard power law with $\alpha= -1.6$
(Halpern \& Wang 1997). 
A parallax distance of 160pc was estimated from HST observations
(Caraveo et al. 1996). \\
{\bf B1055--52} (J1059--5237)\ 
Analyzing BeppoSAX data, Mineo et al.~(2002)
reported that the X-ray spectrum consists of two components:
a soft blackbody with $kT_{\rm s}=78$ eV and 
$A_{\rm s}= 1.3 A_* (d/1.53)^2$
and a hard blackbody with $kT_{\rm h}=740$ eV and 
$A_{\rm h}= 4.8 \times 10^{-6} (d/1.53)^2$.
They pointed out that a blackbody$+$power-law model
also fits the data.
However, the results differ little between the two models, 
by virtue of the negative feedback effect
of the gap electrodynamics (\S~\ref{sec:stability}).
Thus, we adopt the former, two-blackbody model.
The distance is estimated to be $1.53$~kpc from
dispersion measure (Taylor \& Cordes 1993).\\

\subsection{The Vela pulsar}
\label{sec:app_vela}
\subsubsection{Acceleration Field and Characteristics}
\label{sec:Ell_char}
We apply the theory to the Vela pulsar.
Let us first consider the spatial distribution of $\Ell$.
For this pulsar, a small created current density
$j_{\rm gap}=4.6 \times 10^{-5}$ gives the best-fit spectrum
(see \S~\ref{sec:powerlaw} for details). 

\begin{figure} 
\centerline{ \epsfxsize=8.5cm \epsfbox[200 20 470 200]
              {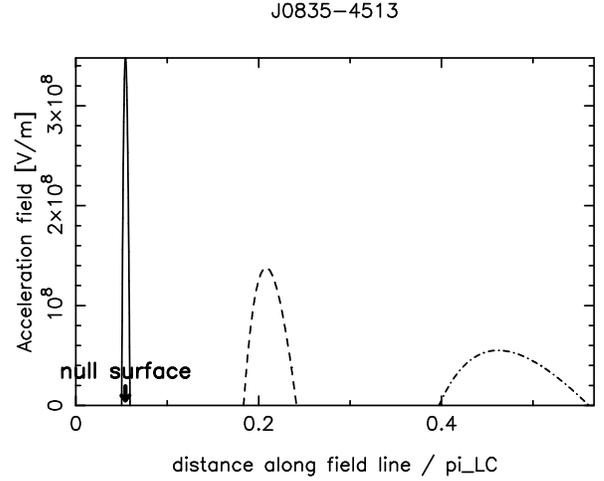} } 
\caption{
Spatial distribution of $\Ell(s)$ 
for $j^{\rm in}=0$ (solid), $0.25$ (dashed), 
and $0.5$ (dash-dotted),
for the Vela pulsar when $\inc= 75^\circ$
and $j_{\rm gap}=4.6 \times 10^{-5}$ and $j^{\rm out}=0$.
The abscissa designates $s/\rlc$,
the distance along the last-open field line
normalized by the light cylinder radius.
        }
\label{fig:EVela_75z} 
\end{figure}

To compare the effects of particle injection,
we present the $\Ell$ distribution
for the three cases of 
$j^{\rm in}=0$ (solid), 0.25 (dashed), and 0.50 (dash-dotted)
in figure~\ref{fig:EVela_75z}.
The magnetic inclination is chosen to be $\inc=75^\circ$.
We adopt $j^{\rm out}=0$ throughout this paper,
unless its value is explicitly specified.

As the solid line shows, the gap is located around the null surface
when there is no particle injection across either of the boundaries.
Moreover, $\Ell$ varies quadratically, because the Goldreich-Julian
charge density deviates from zero linearly near to the
null surface.

As the dashed and dash-dotted lines indicate,
the gap shifts outwards as $j^{\rm in}$ increases.
When $j^{\rm in}=0.5$ for instance, 
the gap is located on the half way
between the null surface and the light cylinder.
This result is consistent with the analytic prediction given in 
\S~\ref{sec:null}.
The gap width increases as it shifts outwards,
because $\lambda_{\rm p}$ in equation~(\ref{eq:closure})
increases due to decreased X-ray density at large distances from the star.

\begin{figure} 
\centerline{ \epsfxsize=8.5cm \epsfbox[200 20 470 200]
              {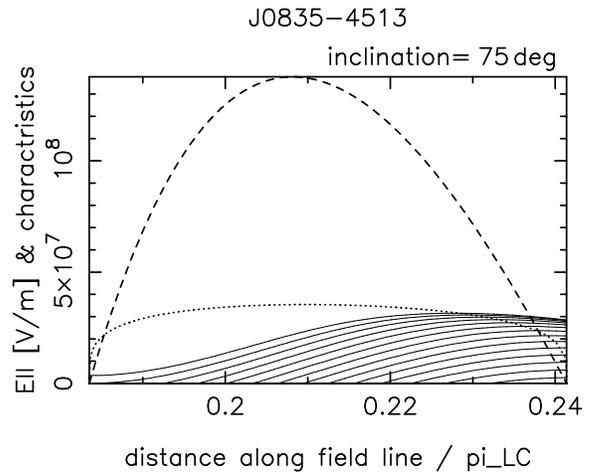} } 
\caption{
Spatial distribution of $\Ell(s)$ (dashed)
for the Vela pulsar when $\inc= 75^\circ$,
$j^{\rm in}=0.25$, $j^{\rm out}=0$,
and $j_{\rm gap}= 4.6 \times 10^{-5}$.
The characteristics for positrons are also shown by solid lines.
The equilibrium Lorentz factor, which would be obtained
if the curvature radiation-reaction force balanced
the electrostatic acceleration, is indicated by the dotted line.
        }
\label{fig:EVela_75a} 
\end{figure}

In figure~\ref{fig:EVela_75a},
we present the characteristics of 
partial differential equation~(\ref{eq:BASIC_2a})
for positrons by solid lines, together with 
$\Ell(s)$ when $j^{\rm in}=0.25$ and $\inc=75^\circ$
(i.e., the dashed line in fig.~\ref{fig:EVela_75z}).
We also superpose the equilibrium Lorentz factor 
that would be obtained
if we assumed the balance between the curvature radiation reaction
and the electrostatic acceleration, as the dotted line.
It follows that the particles are not saturated at the
equilibrium Lorentz factor in most portions of the gap.

In the outer part of the gap where $\Ell$ is decreasing,
characteristics begin to concentrate; as a result, 
the energy distribution of outwardly propagating particles
forms a \lq shock' in the Lorentz factor direction.
However, the particle Lorentz factors do not match 
the equilibrium value (dotted line).
For example, near the outer boundary,
the particles have larger Lorentz factors compared with the
equilibrium value, 
because the curvature cooling scale is longer than the gap width.
Thus, we must discard the mono-energetic approximation 
that all the particles migrate at the equilibrium Lorentz factor
as adopted in Papers~I through IX.
We instead have to solve the energy dependence of the particle 
distribution functions explicitly.

The particles emit $\gamma$-rays not only inside of the gap
but also outside of it,
being decelerated by the curvature radiation-reaction force.
The length scale of the deceleration is given by
\begin{eqnarray}
  l_{\rm curv} 
  &=& c \cdot \frac{\Gamma m_{\rm e}c^2}
                   {\displaystyle\frac{2e^2}{3c^3}\Gamma^4
                           \left(\frac{c^2}{\rho_{\rm c}}\right)^2}
  \nonumber\\
  &=& 0.4 \rlc \Omega_2{}^{-1}
      \left(\frac{\Gamma}{10^7}\right)^{-3}
      \left(\frac{\rho_{\rm c}}{\rlc/2}\right)^2
  \label{eq:cool_curv}
\end{eqnarray}
Since the typical Lorentz factor is a few times of $10^7$,
$l_{\rm curv}$ is typically much less than $\rlc$.
Therefore, the escaping particles lose most of their energies 
well inside of the light cylinder.


\subsubsection{Particle Energy Distribution}
\label{sec:spc_particle}
\begin{figure} 
\centerline{ \epsfxsize=8.5cm \epsfbox[200 20 470 200]
              {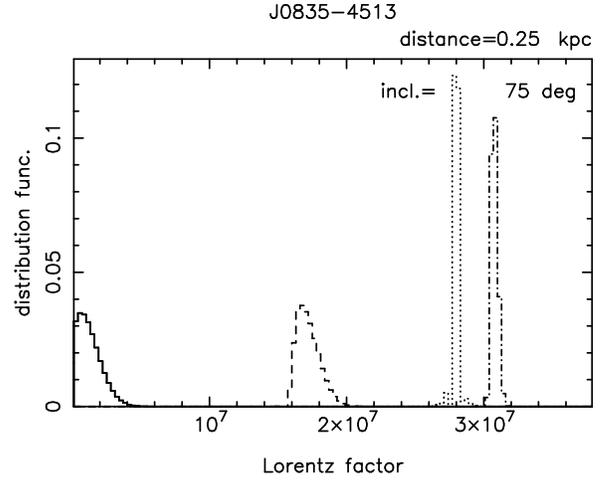} } 
\caption{
Particle energy distribution at several points along the magnetic field
lines for the same case as in figure~\ref{fig:EVela_75a}.
Initial spectrum (solid line) 
evolves to dashed, dash-dotted, and dotted lines,
as positrons propagate outwards. 
}
\label{fig:PVela_75a} 
\end{figure}

As we have seen in the foregoing subsection,
the distribution function of the particles
forms a \lq shock' in the Lorentz factor direction.
In figure~\ref{fig:PVela_75a},
we present the energy distribution of positrons at several 
representative points along the field line.
At the inner boundary ($s=0.184 \rlc$),
particles are injected with Lorentz factors
typically less than $4\times 10^6$ 
as indicated by the solid line.
Particles migrate along the characteristics in the phase space
and gradually form a \lq shock' 
as the dashed line (at $s=0.205\rlc$) indicates,
and attains maximum Lorentz factor 
at $s=0.228\rlc$ as the dash-dotted line indicates.
Then they begin to be decelerated gradually
and escape from the gap
with large Lorentz factors $\sim 2.8 \times 10^7$ (dotted line)
at the outer boundary, $s=s^{\rm out}=0.241\rlc$.

Even though the \lq shock' is captured by only a few  
grid points for the dash-dotted line in figure~\ref{fig:PVela_75a},
the CIP scheme accurately conserves the total current density,
\begin{eqnarray}
  j_{\rm tot}
  &=& \frac{2\pi ce}{\Omega B(s)}
      \int_1^\infty \left[ N_+(s,\Gamma)+N_-(s,\Gamma) \right] d\Gamma
  \nonumber\\
  &=& j^{\rm in}+j^{\rm out}+j_{\rm gap} \approx j^{\rm in}.
  \label{eq:consv_3}
\end{eqnarray}
For this case,
$j_{\rm tot}$ is accurately conserved at $0.25$ level
within 0.2\% errors even at the \lq shock'.

\subsubsection{Formation of Power-law Gamma-ray Spectrum}
\label{sec:powerlaw}
So far, we have seen that the outwardly propagating particles 
are not saturated at the equilibrium value
and that such particles escape from the gap
with sufficient Lorentz factors suffering subsequent cooling 
via curvature process.
It seems, therefore, reasonable to suppose that
a significant fraction of the $\gamma$-ray luminosity 
is emitted from such escaping particles.

\begin{figure} 
\centerline{ \epsfxsize=8.5cm \epsfbox[200 20 470 200]
              {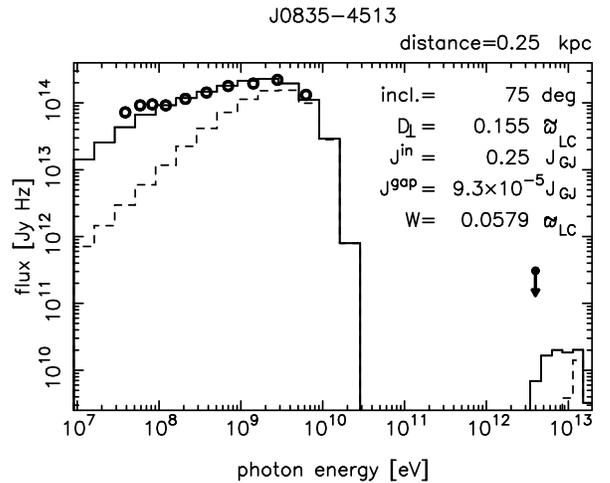} } 
\caption{
Computed $\gamma$-ray spectrum for the Vela pulsar
for the same case as in figure~\ref{fig:EVela_75a}.
The dashed line depicts the flux emitted within the gap,
while the solid one includes that emitted outside it.
        }
\label{fig:SVela_75b}
\end{figure}

We present in figure~\ref{fig:SVela_75b} 
the $\gamma$-ray spectrum 
emitted from outwardly propagating particles (i.e., positrons)
for the same case as in figure~\ref{fig:EVela_75a}.
The dashed line represents the $\gamma$-ray flux emitted
within the gap,
while the solid one includes that emitted outside of the gap
by the escaping particles.
Therefore, the difference between the solid and the dashed
lines indicates the $\gamma$-ray flux emitted 
by the particles migrating outside of the gap.
For comparison, we plot the phase-averaged 
EGRET spectrum, which is approximated by a
power law with a photon index $-1.7$ (Kanbach et al.~1994)
by open circles.

It follows from the figure that the $\gamma$-ray spectrum 
in 100~MeV--3~GeV energies can be explained by the 
curvature radiation emitted by the escaping particles.
We adjusted the transfield thickness as
$D_\perp=0.16\rlc=2.8W$ so that the observed flux may be explained.
The luminosity of the $\gamma$-rays emitted outside of the gap
contribute 48\% of the total luminosity 
$5.08 \times 10^{33} \mbox{ergs s}^{-1}$
between 100~MeV and 20~GeV.
In another word, we do not have to assume a power-law energy
distribution for particles 
(as assumed in some of the previous outer-gap models)
to explain the power-law $\gamma$-ray spectrum for the Vela pulsar.
This conclusion is natural, 
because a power-law energy distribution of particles will not be 
achieved by an electrostatic acceleration,
and because magnetohydrodynamic shocks (i.e., real shocks)
will not be formed in the accelerator.

Because the X-ray field is dense for this young pulsar,
the pair-production mean free path, 
and hence the gap width becomes small
(for details, see 
 Hirotani \& Okamoto 1998; Papers~IV \& V).
As a result, the potential drop in the gap 
$2.24 \times 10^{13}$~V is only 0.81~\% of the 
electro-motive force (EMF) exerted on the spinning neutron star surface
$\sim \mu/\rlc^2 = 2.79 \times 10^{15}$~V.
Nevertheless, this potential drop is enough
to accelerate particles into high Lorentz factors, $10^{7.5}$.

\subsubsection{Solutions in a Wide Parameter Space}
\label{sec:Vela_param}
With the hydrogen atmosphere model,
we can explain the observed $\gamma$-ray spectrum 
in a wide parameter space
$45^\circ < \inc < 75^\circ$ and 
$0.125 < j^{\rm in} < 0.25$,
by appropriately choosing $j_{\rm gap}$ and $D_\perp$.
With increasing $j_{\rm gap}$ ($\ll 1$),
the $\nu F_\nu$ [Jy~Hz] peak energy increases 
because of the increased $W$, 
while the sub-GeV spectrum becomes hard
because of the significant $\gamma$-ray emission within the extended gap 
(rather than outside of it).
On the other hand, $D_\perp$ affects only 
the normalization of the $\gamma$-ray flux;
the $\gamma$-ray luminosity is proportional to $D_\perp^2$.

Let us first fix $j^{\rm in}$ at $0.25$ and consider 
how the best-fit values of $j_{\rm gap}$ and $D_\perp$
depend on $\inc$.
As we have seen, they are $j_{\rm gap}=4.6 \times 10^{-5}$ and 
$D_\perp=0.155\rlc=0.84 s^{\rm in}$ for $\inc=75^\circ$. 
However, the ratio $D_\perp / s^{\rm in}$ 
increases with decreasing $\inc$ and becomes $1.06$ for $45^\circ$.
From geometrical consideration, 
we conjecture that $D_\perp$ should not greatly exceed $s^{\rm in}$;
we thus consider that $\inc > 45^\circ$ is appropriate 
for $j^{\rm in}=0.25$.
A large $\inc$ is preferable to obtain a small $D_\perp / \rlc$.
However, since the radio pulsation shows a single peak, 
we consider that $\inc$ is not close to $90^\circ$.
On these grounds,
we adopted $\inc=75^\circ$ as a compromise for the Vela pulsar.

Let us next fix $\inc$ at $75^\circ$ and 
consider how the best-fit values of $j_{\rm gap}$ and $D_\perp$
depend on $j^{\rm in}$. 
The ratio $D_\perp / \rlc$ increases with decreasing $j^{\rm in}$
and becomes $0.94$ for $j^{\rm in}=0.125$
(c.f. $0.84$ for $j^{\rm in}=0.25$).
This is because the decreased flux of the outwardly migrating particles
(due to decreased $j^{\rm in}$) must be compensated by a large $D_\perp$
to produce the same $\gamma$-ray flux.
Thus, we consider $j^{\rm in}>0.125$ is appropriate for $\inc=75^\circ$.
A large $j^{\rm in}$ is preferable to obtain a small $D_\perp / \rlc$.
However, for $j^{\rm in}>0.25$,
the gap is so extended that 
a significant $\gamma$-rays are emitted above GeV within the gap;
as a result, the sub-GeV spectrum becomes too hard.
On these grounds, we adopted $j^{\rm in}=0.25$ as a compromize.


For $45^\circ < \inc < 75^\circ$,
$0.125 < j^{\rm in}<0.25$,
and appropriately chosen $j_{\rm gap}$ and $D_\perp$,
TeV flux is always less than $3 \times 10^{10}$~JyHz.
Thus, one general point becomes clear:
TeV flux is unobservable with current ground-based telescopes,
provided that the emission solid angle is 1~ster
and that the surface thermal (not magnetospheric) X-rays 
are upscattered inside and outside of the gap.
Since the magnetospheric X-rays will be beamed away from the
gap and their specific intensity is highly uncertain,
we leave the problem of the upscatterings of magnetospheric 
(power-law) X-rays untouched.

\subsection{PSR~B1706-44}
\label{sec:app_1706}
We next apply the theory to a Vela-type pulsar, PSR~B1706-44.
To consider $D_\perp / s^{\rm in}$ as small as possible,
we adopt a large magnetic inclination $75^\circ$.
We compare $\nu F_\nu$ spectra for the three cases:
$j^{\rm in}=0.4$, $0.2$, and $0.1$.

\begin{figure} 
\centerline{ \epsfxsize=8.5cm \epsfbox[200 20 470 200]
              {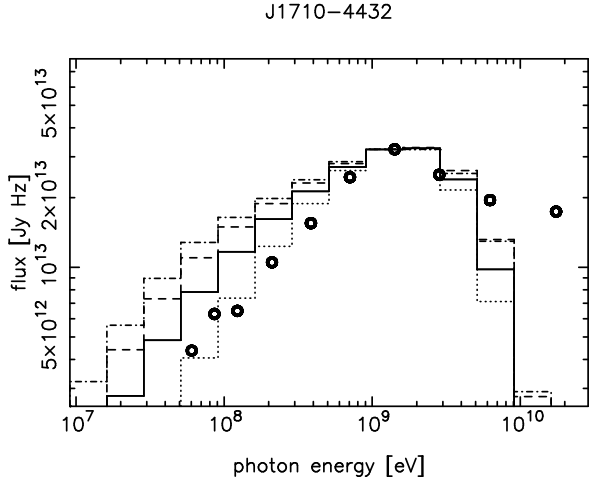} } 
\caption{
Computed $\gamma$-ray spectra for PSR~B1706-44.
The solid, dashed, dash-dotted lines
represent the spectra 
for $j^{\rm in}=0.4$, $0.2$, $0.1$, respectively, 
when $d=2.5$~kpc and $\inc=75^\circ$, 
while the dotted one for 
$j^{\rm in}=0.4$ when $d=1.0$~kpc and $\inc=45^\circ$.
        }
\label{fig:S1706} 
\end{figure}

To examine how the sub-GeV spectrum depends on $j^{\rm in}$, 
we fix the $\nu F_\nu$ peak at the observed value, $\sim 2$~GeV,
by adjusting $j_{\rm gap}$ appropriately.
For the solid ($j^{\rm in}=0.4$), 
dashed ($j^{\rm in}=0.2$), and 
dash-dotted ($j^{\rm in}=0.1$) lines 
in figure~\ref{fig:S1706}, we adopt
$j_{\rm gap}=2.2 \times 10^{-4}$, $1.8 \times 10^{-4}$, and 
$1.5 \times 10^{-4}$, respectively, and $d=2.5$~kpc.
Moreover, the perpendicular thickness is adjusted so that
the predicted flux may match the observed value 
($3.2 \times 10^{13}$~JyHz) at $1.4$~GeV;
for the solid, dashed, and dash-dotted lines, they are
$D_\perp=1.05\rlc=3.5s^{\rm in}$, 
$0.61\rlc=4.1s^{\rm in}$, $0.49\rlc=5.3s^{\rm in}$, respectively.
The sub-GeV spectrum becomes hard with increasing $j^{\rm in}$,
because the ratio of the flux emitted outside of the gap
and that emitted within it
decreases as the gap extends with increasing $j^{\rm in}$.
However, as the solid line indicates, 
the obtained sub-GeV spectrum is still too soft to match the observation.
For a smaller $j^{\rm in}$ ($<0.1$), 
the sub-GeV spectrum becomes further soft.
For a larger $j^{\rm in}(>0.4)$, $D_\perp$ exceeds $\rlc$.
For a non-zero $j^{\rm out}$, 
the inwardly shifted gap is shrunk to emit smaller $\gamma$-ray flux, 
which results in a further greater $D_\perp/s^{\rm in}$.
On these grounds, the predicted sub-GeV spectrum becomes too soft 
or the $\gamma$-ray flux becomes too small 
(i.e., $D_\perp$ becomes too large) for any parameter set
of $j^{\rm in}$ and $j^{\rm out}$, if $\inc=75^\circ$ and $d=2.5$~kpc.

For a small inclination ($\inc<75^\circ$),
the gap is located relatively outside of the magnetosphere,
because the null surface crosses the last-open field line
at large distances from the star.
Since $\lambda_{\rm p}$ increases in equation~(\ref{eq:closure})
at large distances from the star, 
the gap is extended for a small magnetic inclination.
In such an extended gap, 
particles saturate at the equilibrium Lorentz factor
in the outer part (Takata et al.~2002)
and emit most of the $\gamma$-rays around the 
central energy of curvature radiation.
As a result, a hard sub-GeV spectrum can be expected;
however, we have to assume a large $D_\perp$
that exceeds $\rlc$ if $d=2.5$~kpc.

Nevertheless,
if $d$ is much less than $2.5$~kpc and $\inc<75^\circ$,
we can explain the observed spectrum with moderate $D_\perp$.
For example,
for $d=1$~kpc, $\inc=45^\circ$, $j^{\rm in}=0.4$, and
$j_{\rm gap}=3.8\times 10^{-3}$,
the gap exists in $0.51\rlc<s<0.74\rlc$
and the resultant spectrum (dotted line in figure~\ref{fig:S1706})
matches the observation relatively well
with a marginally acceptable thickness, 
$D_\perp=0.80\rlc=1.5 s^{\rm in}$.

On the other hand, for a large inclination ($\inc>75^\circ$), 
the sub-GeV spectrum becomes softer than $\inc=75^\circ$ case
for the same $j^{\rm in}$, $j^{\rm out}$, and $j_{\rm gap}$.
Therefore, the spectrum will not match the observation 
whatever distance we may assume.

We can alternatively consider a CHR-like outer gap.
Assuming a small $D_\perp$ (say, $0.05\rlc$),
we find that the gap extends along the field lines 
due to the screening effect of the zero-potential walls
(i.e., the first term in the right-hand side of eq.~[\ref{eq:BASIC_1}]).
In most portions of this extended gap,
particles are nearly saturated at the equilibrium Lorentz factor.
As a result, the sub-GeV spectrum becomes hard and match the observation
with appropriate peak energy around $2$~GeV.
However, in this case, the $\gamma$-ray flux becomes too small 
to match the observed value, unless we adopt an unrealistic distance
($300$~pc for $D_\perp=0.05\rlc$).

In short, the phase-averaged EGRET spectrum 
for this pulsar cannot be explained either by our current model
or by the CHR picture ($D_\perp \ll W$) for any combinations of 
$\inc$, $j^{\rm in}$, $j^{\rm out}$, $j_{\rm gap}$, and $D_\perp$,
if $d=2.5$~kpc. 
Therefore, we suggest a small distance (e.g., $1$~kpc) for this pulsar
with a large $D_\perp$ ($\sim \rlc$).

\subsection{The Geminga Pulsar}
\label{sec:app_Gemi}
Let us apply the theory to a cooling neutron star, 
the Geminga pulsar.
For a small $\inc$ (e.g., $45^\circ$), 
the gap is so extended that the outer boundary exceeds the
light cylinder.
For a larger $\inc$, on the other hand,
not only the outer-gap emission, but also a polar-cap one
could be in our line of sight.
Since there has been no radio pulsation confirmed,
we consider a moderate magnetic inclination $\inc=60^\circ$.

\begin{figure} 
\centerline{ \epsfxsize=8.5cm \epsfbox[200 20 470 200]
              {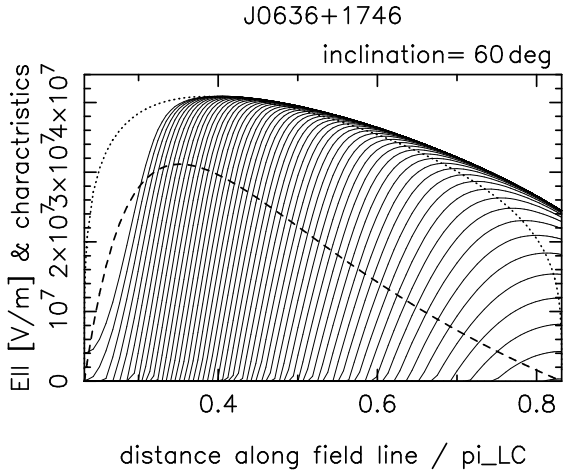} } 
\caption{
Spatial distribution of $\Ell(s)$ (dashed line)
for the Geminga pulsar when $\inc= 60^\circ$, $j^{\rm in}= 0.25$,
$j^{\rm out}=0$, and $j_{\rm gap}=8.0 \times 10^{-5}$.
Particles are saturated at the equilibrium Lorentz factor
(dotted line) for this middle-aged pulsar.
        }
\label{fig:EGemi_60} 
\end{figure}

\begin{figure} 
\centerline{ \epsfxsize=8.5cm \epsfbox[200 20 470 200]
              {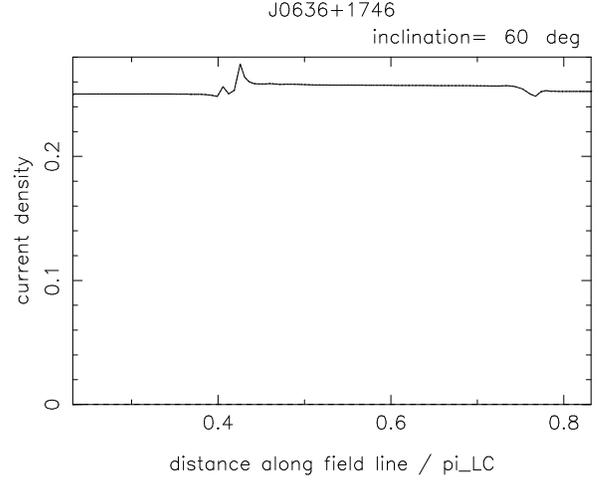} } 
\caption{
Total current density, $j_{\rm tot}$, 
for the same case as in figure~\ref{fig:EGemi_60}.
Even though the particle distribution function forms a strong \lq shock'
in the Lorentz factor direction in $0.4 \rlc < s < 0.45 \rlc$
(see fig.~\ref{fig:EGemi_60}),
$j^{\rm tot}$ is conserved relatively accurately.
        }
\label{fig:CGemi_60} 
\end{figure}

Since the soft photon field is less dense compared with 
young pulsars like Vela or B1706--44,
the gap is extended along the field lines.
For the set of parameters $j^{\rm in}=0.25$, $j^{\rm out}=0$,
and $j_{\rm gap}=8.0\times 10^{-5}$,
which gives the best-fit $\gamma$-ray spectrum,
we obtain $W = 0.59 \rlc$.
We present the spatial distribution of $\Ell$ 
obtained for this set of parameters 
as the dashed line in figure~\ref{fig:EGemi_60},
as well as the characteristics (solid lines).
In the outer part of this extended gap,
$\rhoGJ$, which gradually increases, 
is partially canceled by the 
$-\Psi/D_\perp^2$ term in equation~(\ref{eq:BASIC_1}).
As a result, $\Ell(s)$ deviates from quadratic distribution
and decrease gradually as well.
Because of this extended structure,
particles are nearly saturated at the equilibrium Lorentz factor
(dotted line in fig.~\ref{fig:EGemi_60}).
In another word, the mono-energetic approximation adopted in
Papers~I--IX is justified for this middle-aged pulsar.
Particle distribution function forms a strong \lq shock',
which is captured only with one grid point,
in $0.4 < s/\rlc < 0.45$.
As a result,
$j_{\rm tot}$ fluctuates a little 
as figure~\ref{fig:CGemi_60} indicates.
Nevertheless, it returns to the $0.2525$ level,
which is $1$~\% greater than the value it should be ($0.250$),
as the characteristics begin to be less concentrated
beyond $s=0.45 \rlc$.

\begin{figure} 
\centerline{ \epsfxsize=8.5cm \epsfbox[200 20 470 200]
              {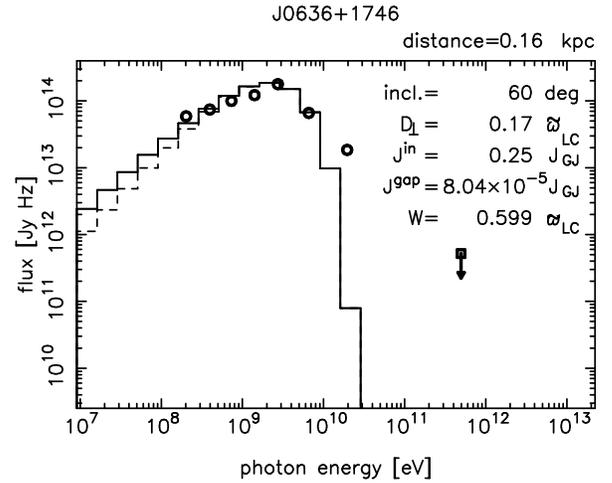} } 
\caption{
Computed $\gamma$-ray spectrum for the Geminga pulsar
for the same case as in figure~\ref{fig:EGemi_60}.
The dashed and solid lines represent the same
components as figure~\ref{fig:SVela_75b}.
        }
\label{fig:SGemi_60} 
\end{figure} 

\begin{figure} 
\centerline{ \epsfxsize=8.5cm \epsfbox[200 20 470 200]
              {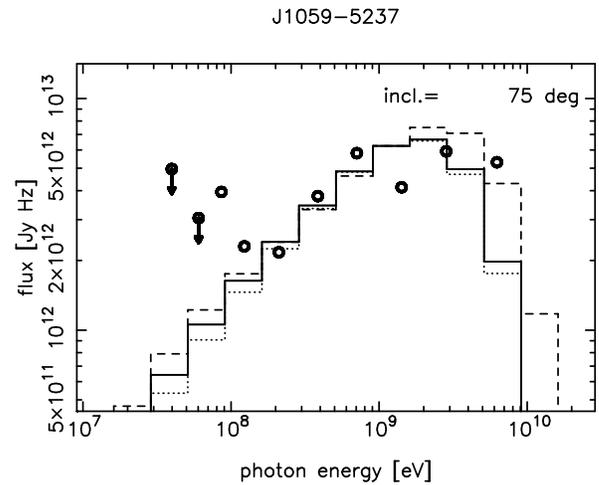} } 
\caption{
Computed $\gamma$-ray spectrum for B1055--52.
Escaping particles little contribute for the luminosity.
For the solid and dashed lines, 
$d=1.53$~kpc is adopted;
the corresponding parameter sets are
$(j^{\rm in},j_{\rm gap},D_\perp/\rlc)$=
$(0.25, 2.9\times 10^{-3},0.27)$, and 
$(0.125,3.7\times 10^{-3},0.20)$, respectively.
For the dotted line,
$d=1$~kpc and 
$(j^{\rm in},j_{\rm gap},D_\perp/\rlc)$=
$(0.25, 4.5\times 10^{-3},0.16)$.
        }
\label{fig:S1055_75} 
\end{figure} 

In figure~\ref{fig:SGemi_60}, we present the resultant $\gamma$-ray
spectrum.
Because of the nearly saturated motion of the particles,
they lose most of their energy within the gap.
As a result,
$\gamma$-ray luminosity associated with the escaping particles 
($5.3 \times 10^{31} \mbox{ergs s}^{-1}$), is
negligibly small compare to that emitted within the gap
($1.11 \times 10^{33} \mbox{ergs s}^{-1}$),
which is represented by the dashed line.

It should be noted that a small 
$D_\perp=0.17\rlc=0.28W$ gives the best-fit spectrum.
This perpendicularly thin gap indicates that the solution
presented in fig.~\ref{fig:EGemi_60}, \ref{fig:CGemi_60}, and
\ref{fig:SGemi_60} are, in fact, obtained in the CHR picture
($D_\perp\ll W$) in the sense that the screening effect due to
the zero-potential walls 
(i.e., $-\Psi/D_\perp^2$ term in eq.~[\ref{eq:BASIC_1}])
is important.
There are, of course, differences from previous works:
In the present work, the set of Maxwell and Boltzmann equations are 
solved, while the two-dimensional screening effects are considered 
only approximately.
To consider the two-dimensional effect more rigorously,
we have to solve the elliptic-type partial differential 
equation~(\ref{eq:Poisson_2D}) on the poloidal plane,
together with ordinary differential equations~(\ref{eq:BASIC_3})
and hyperbolic-type partial differential equations~(\ref{eq:BASIC_2a})
and (\ref{eq:BASIC_2b}) simultaneously.

It follows from figure~\ref{fig:SGemi_60} 
that we can explain the observed spectrum between 200~MeV and 6~GeV 
by superposing the spectra of $\gamma$-rays emitted at various point.
It is interesting to compare this result
with what obtained for the Vela pulsar (fig.~\ref{fig:SVela_75b}).
Between $100$~MeV and $1$~GeV energies,
both spectra are formed by the superposition of the curvature radiation
emitted by the particles having different energies at different
positions.
The important difference is that the particles are saturated at the
equilibrium Lorentz factor {\it in the gap} for the Geminga pulsar,
while they are nearly mono-energetic but only decelerated 
via curvature process {\it outside of the gap} for the Vela pulsar.
Because the particles are no longer accelerated outside of the gap,
they emit $\gamma$-rays in lower energies 
compared with those still being accelerated in the gap.
As a result, the $\gamma$-ray spectrum for the Vela pulsar becomes
softer than that for the Geminga pulsar.
Extending this consideration,
we can predict that a $\gamma$-ray spectrum below GeV
is soft for a young pulsar 
and tends to become hard as the pulsar ages.


\subsection{PSR~B1055-52}
\label{sec:app_1055}
Let us finally apply the present theory to another middle-aged pulsar,
B1055-52.
To obtain a large $\gamma$-ray flux for an appropriately chosen set of
$j^{\rm in}$, $j_{\rm gap}$, and $D_\perp(<s^{\rm in})$,
we adopt a large magnetic inclination, $\inc=75^\circ$.

Since the acceleration field and the particle energy distributions
are similar to the Geminga pulsar, we present only the 
computed $\gamma$-ray spectra for this pulsar 
in figure~\ref{fig:S1055_75}. 
The solid and dashed lines represent the spectra for
the $j^{\rm in}=0.25$ and $0.125$, respectively.
For $j^{\rm in}=0.25$, $j_{\rm gap}= 2.9 \times 10^{-3}$ 
and $D_\perp=0.27\rlc=0.98W$ are chosen 
so that the peak energy of curvature radiation
may match the observed peak energy.
In this case, the gap exists in $0.1485\rlc < s < 0.4120\rlc$.
For $j^{\rm in}=0.125$,
$j_{\rm gap}=3.7 \times 10^{-3}$ and
$D_\perp=0.20\rlc=1.7W$ are chosen
and the gap exists in $0.0870\rlc < s < 0.201\rlc$.
It is interesting to note that the spectrum softens for a
smaller $D_\perp/W$.
This is because the zero-potential wall partially screens $\Ell$
in equation~(\ref{eq:BASIC_1}).
For the middle-aged pulsars Geminga and B1055-52,
the $\gamma$-ray luminosity is not simply proportional to $D_\perp^2$,
because of this screening effect.

It follows from the figure that the solid line 
matches the observed flux with an {\it unreasonable} 
transfield thickness,
$D_\perp=1.8 s^{\rm in}$.
For the dashed line, 
we have to choose $D_\perp=2.3s^{\rm in}$.
The observed fluxes cannot be explained with acceptable
gap width (e.g., $D_\perp < s^{\rm in}$) 
no matter what we may adjust
$j^{\rm in}$, $j^{\rm out}$, $j_{\rm gap}$, and $D_\perp$
if $d=1.53$~kpc.

On these grounds, 
we conjecture that the distance $1.5\pm 0.4$~kpc
determined from the dispersion measure (Taylor \& Cordes 1993)
is too large 
and that a more closer distance, such as
$500$~pc derived from ROSAT data analysis 
($\ddot{\rm O}$gelman \& Finley 1993)
or $700$~pc estimated from a study of the extended nonthermal
radio source around the pulsar 
(Combi, Romero, Azc$\acute{\rm a}$rate 1997), 
is plausible.
For example, if we set $d=1$~kpc, 
we can fit the spectrum with a reasonable transverse thickness,
$D_\perp=1.06s^{\rm in}=0.15\rlc=0.45W$,
for $j^{\rm in}=0.25$ and $j_{\rm gap}=4.5\times 10^{-3}$
(dotted line in fig.~\ref{fig:S1055_75}).  
For such a transversely thin ($D_\perp=0.45W$) gap,
$\Ell$ is significantly screened by the zero-potential wall 
to become CHR-like distribution, 
as in the case for the Geminga pulsar.

In Papers~IX and X,
we suggested a smaller distance, $d<0.5$~kpc, 
using ROSAT and ASCA data (Greiveldinger et al.~1996),
which gives an enormously large blackbody area
$A_{\rm s}=7.3A_\ast(d/1.53\mbox{kpc})^2$.
However, a more recent BeppoSAX data suggest a reasonable
blackbody area of $A_{\rm s}=1.3A_\ast$;
as a result, a larger distance $d<1$~kpc becomes acceptable
in the present analysis.

\section{Discussion}
\label{sec:discussion}
In summary, we have quantitatively examined the stationary 
pair-production cascade in an outer magnetosphere,
by solving the set of Maxwell and Boltzmann equations 
one-dimensionally along the magnetic field lines.
We revealed that an accelerator (or a potential gap) 
is quenched by the created pairs in the gap
but is {\it not} quenched by the injected particles from outside
of the gap,
and that the gap position shifts as a function of the
injected particle fluxes:
If the injection rate across the inner (or outer) 
boundary approaches the typical Goldreich-Julian value,
the gap is located near to the light cylinder (or the star surface).
It should be emphasized that the particle energy distribution
is not represented by a power law,
as assumed in some of previous outer-gap models.
The particles escape from the gap with sufficient Lorentz factors
and emit significant photons in 100~MeV--3~GeV energies
via curvature radiation outside of the gap. 
The $\gamma$-ray spectrum including this component
explains the phase-averaged EGRET spectra for the Vela pulsar 
(and also for PSR~B1706-44 with a small distance, $d<1$~kpc).
As a pulsar ages, its outer gap extends in the magnetosphere
to approach the CHR solution:
For the two middle-aged pulsars Geminga and B1055-52, 
we obtain harder spectra compared with 
the two young pulsars.
TeV fluxes are unobservable with current ground-based telescopes
for all the four pulsars.

We consider the stability of such a gap in the next subsection.
We then point out an implication to the $\gamma$-ray luminosity
versus the spin-down luminosity in \S~\ref{sec:lumin},
and discuss future extensions of the present method
in \S\S~\ref{sec:return}--\ref{sec:unif_pol}.

\subsection{Stability of the Gap}
\label{sec:stability}
Let us discuss the electrodynamic stability of the gap,
by considering whether an initial perturbation of 
the gap width $W$ grows or not.
Imagine that $W$ increases perturbatively.
Then the Maxwell equation~(\ref{eq:Poisson_2D}) increases
$\vert\Ell\vert \equiv -\partial \Psi / \partial s$.
To see this behavior more clearly, we can Taylor expand the
right-hand side around the point $s=s_0$ 
where $\rho_{\rm e}-\rhoGJ$ vanishes.
Neglecting $z$ dependence, we obtain
\begin{equation}
  -\frac{d^2 \Psi}{d s^2}= -4\pi A (s-s_0);
  \label{eq:Poisson_vacuum}
\end{equation}
the constant $A$ is of the order of $\Omega\mu_{\rm m}/(c\rlc^4)$, 
where $\mu_{\rm m}$ is the magnetic dipole moment of the star
(for an explicit expression of $A$, see eq.~[3] in Paper~V).
If $W \ll \rlc$, $s_0$ approaches $s_{\rm cnt}$ (eq.~[\ref{eq:def_scnt}]).
Integrating equation~(\ref{eq:Poisson_vacuum}), 
we obtain $\Ell(s)=2\pi A (s^{\rm out}-s)(s-s^{\rm in})$,
where $s^{\rm out}= s^{\rm in}+W$.
Thus, the averaged $\Ell$ in the gap can be evaluated as
\begin{equation}
  \langle{\Ell}\rangle
  = \frac{1}{W}\int_{s^{\rm in}}^{s^{\rm out}} \Ell(s) ds
  = \frac{\pi}{3} A W^2
  \label{eq:ave_Ell}
\end{equation}

Particle Lorentz factor $\Gamma$ also increases with increasing $W$.
For example, if particles are unsaturated, 
particles energy is roughly proportional to the potential drop;
thus, we obtain 
\begin{equation}
  \Gamma \sim \langle\Ell\rangle \cdot W \propto W^3.
  \label{eq:Lf_1}
\end{equation}
On the contrary, if they are saturated,
the curvature radiation drag gives
\begin{equation}
  \Gamma \sim \left( \frac{3\rho_{\rm c}\langle\Ell\rangle}{2e} 
              \right)^{1/4}
  \propto W^{1/2}
  \label{eq:Lf_2}
\end{equation}
In general, $\Gamma$ depends on $W^\alpha$ with $0.5<\alpha<3$;
thus, $\Gamma$ increases with increasing $W$ 
irrespectively whether the particles are saturated or not.

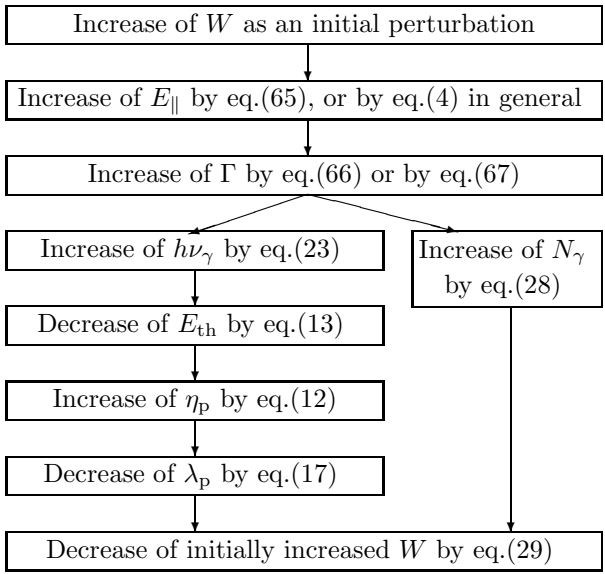
\begin{figure}
  \setlength{\unitlength}{1mm}
  \begin{picture}(79,150)(0,70)
    \put( 0,140){\framebox(79,5){\shortstack{%
       Increase of $W$ as an initial perturbation}}}
    \put(40,140){\vector(0,-1){5}}
    \put( 0,130){\framebox(79,5){\shortstack{%
       Increase of $\Ell$ 
       by eq.(\ref{eq:ave_Ell})}, or 
       by eq.(\ref{eq:Poisson_2D}) in general }}
    \put(40,130){\vector(0,-1){5}}
    \put( 0,120){\framebox(79,5){\shortstack{%
       Increase of $\Gamma$
       by eq.(\ref{eq:Lf_1}) or
       by eq.(\ref{eq:Lf_2})    }}}
    \put(40,120){\vector(-3,-1){15.8113}}
    \put( 0,110){\framebox(50,5){\shortstack{%
       Increase of $h\nu_\gamma$ 
       by eq.(\ref{eq:def_Ec})} }}
    \put(25,110){\vector(0,-1){5}}
    \put( 0,100){\framebox(50,5){\shortstack{%
       Decrease of $E_{\rm th}$
       by eq.(\ref{eq:def_Eth})} }}
    \put(25,100){\vector(0,-1){5}}
    \put( 0, 90){\framebox(50,5){\shortstack{%
       Increase of $\etaP$
       by eq.(\ref{eq:def_etap_0}) }}}
    \put(25, 90){\vector(0,-1){5}}
    \put( 0, 80){\framebox(50,5){\shortstack{%
       Decrease of $\lambda_{\rm p}$
       by eq.(\ref{eq:def_mfp})} }}
    \put(25, 80){\vector(0,-1){5}}
    \put(40,120){\vector(+4,-1){20.000}}
    \put(54,105){\framebox(25,10){\shortstack{%
       Increase of $N_\gamma$ \\
       by eq.(\ref{eq:Ngamma})} }}
    \put(67,105){\vector(0,-1){30}}
    \put( 0, 70){\framebox(79,5){\shortstack{%
       Decrease of initially increased $W$
       by eq.(\ref{eq:closure})} }}
  \end{picture}
  \caption{Stability of an outer-gap accelerator}
  \label{fig:stability} 
\end{figure}

As a result of this increased $\Gamma$, 
the $\gamma$-ray energy, $h\nu_\gamma$, increases.
This is because the central energy of the curvature radiation increases
with increasing $\Gamma$ (eq.~[\ref{eq:def_Ec}]).
Moreover, the number of $\gamma$-rays emitted by a single particle,
$N_\gamma \propto W \cdot \Gamma$, increases with
increasing $W$ (eq.~[\ref{eq:def-etaC}] \& [\ref{eq:Ngamma}]).
The flowchart is depicted in figure~\ref{fig:stability}.

The increased $h\nu_\gamma$ results in a decrease of 
$E_{\rm th}$ by equation~(\ref{eq:def_Eth}).
The decreased $E_{\rm th}$ leads to the increase of
$\etaP$ by equation~(\ref{eq:def_etap_0}), 
and hence to the decrease of $\lambda_{\rm p}$ by 
equation~(\ref{eq:def_mfp}).
This result does not depend on the origin of the X-ray field.
For example, if the X-rays are thermal origin,
$\lambda_{\rm p}$ decreases with 
decreasing threshold energy $E_{\rm th}$,
because the specific intensity of the X-ray field will be unchanged.
If the X-rays are magnetospheric origin, on the other hand,
$\lambda_{\rm p}$ decreases more sharply than the thermal case,
because the X-ray density illuminating the gap will increase
with increasing pair-production rate outside of the gap
due to the increased $h\nu_\gamma$ (and $N_\gamma$).

Finally, both the decreased $\lambda_{\rm p}$ 
and the increased $N_\gamma$
contribute to reduce the initial increase of $W$
by equation~(\ref{eq:closure}).
Because of this negative feedback effect,
stationary solutions exist in a wide range of pulsar parameters
(from young to middle-aged pulsars).
On these grounds,
although the perturbation equations are not solved
under appropriate boundary conditions for the perturbed quantities,
we conjecture that an outer gap is electrodynamically stable,
irrespective whether the X-ray field illuminating the gap
is thermal or non-thermal origin.

\subsection{Gamma-ray vs. Spin-down Luminosities}
\label{sec:lumin}
It should be noted that 
the emission from the escaping particles attain typically 
$40\%$ of the total $\gamma$-ray luminosity for young pulsars.
Thus, it is worth mentioning its relationship 
with the spin-down luminosity,
\begin{equation}
  L_{\rm spin}= -I \Omega \dot{\Omega}  \propto \Omega^{n+1},
  \label{eq:spindown}
\end{equation}
where the braking index $n$ is related to the spin-down rate as
\begin{equation}
  \dot{\Omega}= -k \Omega^n.
  \label{eq:braking}
\end{equation}
If the spin down is due to the magnetic dipole radiation, 
we obtain $n=3$.

The outwardly propagating particles escape from the gap 
with spatial number density 
\begin{equation}
  N^{\rm out}
  = (j^{\rm in}+j_{\rm gap})\frac{\Omega B^{\rm out}}{2\pi ce},
  \label{eq:NeOUT}
\end{equation}
where $B^{\rm out}=B(s^{\rm out})$.
Therefore, the energy carried by the escaping particles 
per unit time is given by
\begin{equation}
  L_{\rm esc}
  = D_\perp^2 c N^{\rm out} \Gamma_{\rm esc} m_{\rm e}c^2,
  \label{eq:def_Lesc}
\end{equation}
where $\Gamma_{\rm esc} (\sim 10^{7.5})$ refers to the Lorentz factor 
of escaping particles.
Note that $\Gamma_{\rm esc}$ is essentially determined by the
equilibrium Lorentz factor (dotted line in fig.~\ref{fig:EVela_75a})
near the gap center.
Since the equilibrium Lorentz factor depends on the one-fourth power
of $\Ell$, the variation of $\Gamma_{\rm esc}$ on pulsar parameters
is small.
We can approximate $B^{\rm out}$ as
\begin{equation}
  B^{\rm out} \sim \frac{\mu_{\rm m}}{\rlc^3}  
     \left( \frac{\rlc}{r^{\rm out}} \right)^3,
  \label{eq:Bout}
\end{equation}
where $r^{\rm out}$ refers to the distance of the outer boundary 
of the gap from the star center.
Let us assume that the position of the gap with respect to 
the light cylinder radius,
$r^{\rm out}/\rlc$, 
does not change as the pulsar evolves; this situation can be realized if 
$j^{\rm in}-j^{\rm out}$ is unchanged.
Evaluating $B$ at $r=0.5\rlc$, we obtain
\begin{eqnarray}
  L_{\rm esc} 
  &=& \frac{4 \Gamma_{\rm esc} m_{\rm e}c}{\pi e} 
      \mu_{\rm m} \Omega^2 \left(\frac{D_\perp}{\rlc}\right)^2
  \nonumber\\
  &\propto& L_{\rm spin}{}^{0.5},
  \label{eq:Lesc}
\end{eqnarray}
where $n=3$ is assumed in the second line.
To derive this conclusion,
it is essential that the particles are not saturated
at the equilibrium Lorentz factor.
Thus, the same discussion can be applied irrespective of the
gap position or the detailed physical processes involved.
For example, an analogous conclusion was derived for a polar-cap model
by Harding, Muslimov, and Zhang (2002).
It is, therefore, concluded that the observed relationship
$L_\gamma \propto L_{\rm spin}{}^{0.5}$ 
merely reflects the fact that the particles are
unsaturated in the gap
and does not discriminate the gap position.

Let us compare this result with what would be expected in 
the CHR picture.
Since the gap is extended significantly along the field lines 
in the CHR picture,
particles are saturated at the equilibrium Lorentz factor
to lose most of their energies within the gap,
rather than after escaping from it.
We can therefore estimate the $\gamma$-ray luminosity as
\begin{equation}
  L_{\rm gap}
  = (D_\perp D_\phi W)
    \cdot N^{\rm out}
    \cdot \Pcv
  \label{eq:def_Lgap}
\end{equation}
where $D_\phi$ refers to the azimuthal thickness of the gap.
Noting that the particle motion saturates at the equilibrium 
Lorentz factor satisfying $\Pcv/c=e\Ell$,
recalling that the acceleration field is given by
$\Ell \approx \Omega B D_\perp^2 / 4\rho_{\rm c}c$
in the CHR picture,
and evaluating $B$ at $r=\rlc$,
we obtain
\begin{eqnarray}
  L_{\rm gap}
  &=& \frac{\mu_{\rm m}^2 \Omega^4}{4\pi c^3}
      \frac{D_\perp^3 D_\phi W}{\rlc^5}
      \left(\frac{\rho_{\rm c}}{0.5\rlc}\right)^{-1}
  \nonumber\\
  &\propto& L_{\rm spin}
  \label{eq:Lgap},
\end{eqnarray}
where $n=3$ is assumed again in the second line.
Even though the escaping particles little contribute 
to the $\gamma$-ray luminosity in the CHR picture, 
it is worth mentioning the work done by 
Crusius--W$\ddot{\rm a}$tzel and Lesch (2002),
who accurately pointed out 
the importance of the escaping particles
in the CHR picture,
when we interpret $L_\gamma \propto L_{\rm spin}^{0.5}$ relation.

As we have seen,
the particles being no longer accelerated
contribute for the $\gamma$-ray luminosity that is proportional 
to $L_{\rm spin}^{0.5}$.
Reminding that the particles migrate with larger Lorentz factors
than the equilibrium value in the outer part of the gap 
(see fig.~\ref{fig:EVela_75a}),
we can expect roughly half of the $\gamma$-ray luminosity
is proportional to $L_{\rm spin}^{0.5}$
(mainly between 100~MeV and 1~GeV),
and the rest of the half to $L_{\rm spin}$
(mainly above 1~GeV).
As a pulsar ages,
its declined surface emission results in a large 
pair-production mean free path, and hence $W$.
Because $\vert \rhoGJ \vert \propto r^{-3}$ becomes small 
in the outer part of such an extended gap,
$\Ell(s)$ deviates from quadratic distribution 
to decline gradually in the outer part (fig.~\ref{fig:EGemi_60}). 
As a result, particles tend to be saturated at the equilibrium value.
On these grounds, we can predict that
the $\gamma$-ray luminosity tends to be proportional to
$L_{\rm spin}$ with age, 
deviating from $L_{\rm spin}^{0.5}$ dependence for young pulsars.
 
In the present paper, we have examined the set of Maxwell and
Boltzmann equations
one-dimensionally both in the configuration and the momentum spaces
(i.e., only $s$ and $\Gamma$ dependences are considered.) 
In the next three sections,
we discuss the extension of the present method 
into higher dimensions

\subsection{Returning Particles}
\label{sec:return}
If we consider the pitch-angle dependence of particle
distribution functions,
we can compute the radiation spectrum
with synchro-curvature formula (Cheng and Zhang 1996).
Moreover, we can also consider the returning motion of particles
inside and outside of the gap.
The returning motion becomes particularly important
when both signs of charge are injected across the boundary.
For example, not only positrons but also electrons could be
injected across the inner boundary from the polar-cap accelerator.
If $\Ell>0$ for instance, the injected electrons return
in the gap.
This returning motion significantly affects the Poisson equation,
if their injection rate is a good fraction of 
the Goldreich-Julian value.

It remains an unsettled issue whether an outer-gap accelerator
resides on the field lines on which a polar-cap accelerator
exists.
To begin with, 
let us consider the case when the plasma flowing 
between the polar cap and the outer-gap accelerator 
is completely charge separated.
Such a situation can be realized, for instance, 
if only positively charged particles are ejected outwardly 
from the polar cap while
there is virtually no electrons ejected inwardly from 
the outer gap.
Neglecting the pair production, 
current conservation law gives the charge density, $\rho_{\rm e}$,
per unit magnetic flux tube as 
\begin{equation}
  \frac{\rho_{\rm e}}{B} \propto \frac{j_{\rm tot}}{v},
  \label{eq:sep_flow}
\end{equation}
where $v$ refers to the particle velocity along the field line,
and $j_{\rm tot}$ the conserved current density per magnetic flux tube.
At each point along the field line,
$\rho_{\rm e}$ should match $\rho_{\rm GJ}$.
If the field line intersects the null surface, 
$\rho_{\rm e}$ must vanish there;
this obviously violates the causality in special relativity.
Therefore, a stationary ejection of a completely charge-separated plasma
from the polar cap
can be realized only along the field lines between 
the magnetic axis and those intersecting the null surface
at the light cylinder.
On these grounds, it was argued that an outer-gap accelerator,
which is formed close to the last-open field line,
may not resides on the same field lines on which 
a polar-cap accelerator resides.
This has been, in fact, the basic idea that an outer gap will
not be quenched, 
because the particles ejected from the polar cap will flow
along the different field lines.
This idea was welcomed in outer-gap models,
because a gap has been considered to be 
quenched if the external particle injection rate
becomes comparable to the Goldreich-Julian value,
which was proved to be incorrect in this paper.

In general, however, the plasmas are not completely charge separated
and consist of both signs of charge (e.g., positrons and electrons).
Such a situation can be realized, for instance, 
if both charges are ejected outwardly from a polar-cap accelerator,
or if positively charged particles are ejected outwardly
from the polar cap while
electrons are ejected inwardly from the outer gap,
or if there is a pair production between the two accelerators.
In these cases, the velocities of both charges
will be adjusted so that both the current conservation and 
$\rho_{\rm e}=\rho_{\rm GJ}$ are satisfied at each point along the
field lines.
Therefore, it seems likely that a polar-cap accelerator and 
an outer-gap accelerator reside on the same field lines.

To examine if there is a stationary plasma flow between the polar cap
and the outer gap,
we must extend the present analysis
into two dimensional momentum space in the sense that
the pitch-angle dependence of 
the particle distribution functions is taken into account
in addition to the Lorentz factor dependence. 
For example, if both charges are ejected from the polar-cap accelerator,
electrons will return in the outer gap,
screening the original acceleration field in the gap,
and violating the original balance of $\rho_{\rm e}=\rhoGJ$
outside of the gap.
Because the returning motion of particles can be treated correctly
if we consider the pitch-angle evolution of the distribution functions,
and because the pair production is already taken into account,
our present method is ideally suited to investigate the plasma flows
and $\Ell$ distribution self-consistently inside and outside of the gap.

\subsection{Unification of Outer-gap Models}
\label{sec:unif_out}
In addition to the extension into a higher dimensional momentum space,
it is also important to extend the present method 
into a two- or three-dimensional configuration space.
In particular, determination of the perpendicular thickness,
$D_\perp$, 
is important to constrain gap activities.
There have been, in fact, some attempts to constrain $D_\perp$
in the CHR picture.
Since $\Ell$ is proportional to $B D_\perp^2$ if $D_\perp \ll W$, 
particles energies, and hence the $\gamma$-ray energies
increase with increasing $D_\perp$ (for a fixed $B$).
Zhang and Cheng (1997) constrained $D_\perp$, by
considering the condition that the $\gamma$-rays
cause photon-photon pair production in the gap.
Subsequently,
Cheng, Ruderman, and Zhang (2000) extended this idea into
three-dimensional magnetosphere and discussed phase-resolved 
$\gamma$-ray spectra for the Crab pulsar.
In addition, 
Romani (1996) discussed the evolution of the $\gamma$-ray emission
efficiency and computed the phase-resolved spectra for the Vela pulsar,
by assuming that $BD_\perp^2$ declines as $r^{-1}$.
However, in these works, screening effects due to 
pair production has not been considered;
thus, the obtained $D_\perp$, as well as the hypothesized gap position
along the magnetic fields, are still uncertain.

On the other hand, in our approach (picture), 
$D_\perp$ is not solved but only adjusted 
so that the $\gamma$-ray flux may match the observations.
Therefore, the question we must consider next is 
to solve such geometrical and electrodynamical
discrepancies between these two pictures.
We can investigate this issue by extending the present method 
into higher spatial dimensions.

Furthermore, close to the last-open field line,
$\rho_{\rm e}$ may fail to match $\rhoGJ$ outside of the gap,
because the emitted $\gamma$-rays soon propagate away from it
and the pair production is expected to be less efficient.
In such a highly charge-starved region in the magnetosphere,
the electromagnetic field may be approximated by the
Deutsch field (Deutsch 1955).
Therefore, the $\gamma$-rays emitted close to the
last-open field line may show a very hard spectrum
as Higgins and Henriksen (1997; 1998) predicted.
To consider this issue further, we have to solve
the set of Maxwell and Boltzmann equations
on the poloidal plane ($s$,$z$) both inside and outside of the gap,
taking account of the deviation of the $\gamma$-rays
from the field lines on which they were originally emitted.

\subsection{Unification of Outer-gap and Polar-cap Models}
\label{sec:unif_pol}
Electrodynamically speaking, the essential difference 
between outer-gap and polar-cap accelerators 
is the value of the optical depth for pair production.
In an outer-gap accelerator, 
pair production takes place via $\gamma$-$\gamma$ collisions 
and its mean-free path is much greater than the light cylinder radius.
Therefore, a pair production cascade takes place gradually in the gap.
In such a gap, $\Ell$ is automatically screened out 
at both the boundaries
by the Goldreich-Julian charge density,
which varies monotonically along the field lines.
For example, if $\Ell$ is positive, 
$B_\zeta$ in equation~(\ref{eq:BASIC_1}) increases outwards:
$d\Ell/ds>0$ in the inner part of the gap, while
$d\Ell/ds<0$ in the outer part.
Thus, we do not have to contrive a mechanism to screen out $\Ell$
at the boundaries.

On the contrary, in a polar-cap accelerator, 
$\Ell$ cannot be screened out by $\rhoGJ$.
Nevertheless, in the vicinity of the star,
the strong magnetic field (e.g., $B \sim 10^{12}$~G)
leads to a magnetic pair production, 
of which mean free path is much less than the star radius.
As a result, 
a pair production avalanche takes place in a limited region, 
which is called as the \lq pair formation front', in the gap 
(Fawley, Arons, \& Sharlemann 1977; 
 Harding \& Muslimov 1998, 2001, 2002;
 Shibata, Miyazaki, Takahara 1998, 2002;
 Harding, Muslimov, Zhang 2002).
In the pair formation front,
a small portion of the particles return to screen out $\Ell$. 
Such a returning motion 
can be self-consistently solved together 
with $\Ell$ by our present method, 
if we implement the magnetic pair production 
and the resonant IC scattering redistribution functions 
in the source terms of the particles' and $\gamma$-rays' 
Boltzmann equations.
We can execute the same advection-phase computation in CIP scheme; 
thus, all we have to do is to add these source terms 
in the non-advection-phase computation, 
which is not very difficult.
Since analogous boundary conditions 
(e.g., $\Ell=0$ for a space-charge limited flow) will be applied,
we expect the present method is also applicable 
to a polar-cap accelerator.
This is an issue to be examined in our subsequent papers.

\par
\vspace{1pc}\par

The author wishes to express his gratitude to
Drs. A.~K.~Harding, S.~Shibata, and K.~S.~Cheng 
for fruitful discussion on theoretical aspects,
and to Drs. K. Shibata and A. Figueroa-Vin{$\bar{\rm a}$}s 
for valuable advice on numerical analysis.

\end{document}